\documentclass{aa}  

\usepackage{graphicx}
\usepackage{txfonts}
\usepackage{lscape}
\usepackage{booktabs}
\usepackage{threeparttablex}
\usepackage{wasysym}
\usepackage{arydshln}
\usepackage[bookmarks=false,colorlinks=true,linkcolor=black,citecolor=cyan,filecolor=black,urlcolor=cyan]{hyperref}
\usepackage{array}

\newcolumntype{C}[1]{>{\centering\arraybackslash}p{#1}}

\begin{document} 

   \title{Testing the disk-corona interplay in radiatively-efficient broad-line AGN}

   \author{R. Arcodia
          \inst{1}
          \and
          A. Merloni \inst{1}
          \and
          K. Nandra \inst{1}
          \and
          G. Ponti \inst{1,2}          
          }

   \institute{Max-Planck-Institut f\"ur extraterrestrische Physik (MPE), Giessenbachstrasse 1, 85748 Garching bei M\"unchen, Germany\\
              \email{arcodia@mpe.mpg.de}
         \and
		INAF-Osservatorio Astronomico di Brera, Via Bianchi 46, I-23807 Merate (LC), Italy
         }

   \date{Received ; accepted }

 
  \abstract{
  The correlation observed between monochromatic 
  X-ray and UV luminosities
  in radiatively-efficient active galactic nuclei (AGN) lacks a clear theoretical explanation despite being used for many applications. Such a correlation, with its small intrinsic scatter and its slope that is smaller than unity in log space, represents the compelling evidence that a mechanism regulating the energetic interaction between the accretion disk and the X-ray corona must be in place. This ensures that going from fainter to brighter sources the coronal emission increases less than the disk emission. We discuss here a self-consistently coupled disk-corona model that can identify this regulating mechanism in terms of modified viscosity prescriptions in the accretion disk. The model predicts a lower fraction of accretion power dissipated in the corona for higher accretion states. We then present a quantitative observational test of the model using a reference sample of broad-line AGN and modeling the disk-corona emission for each source in the $L_X-L_{UV}$ plane. We used the slope, normalization, and scatter of the observed relation to constrain the parameters of the theoretical model. For non-spinning black holes and static coronae, we find that the accretion prescriptions that match the observed slope of the $L_X-L_{UV}$ relation produce X-rays that are too weak with respect to the normalization of the observed relation. Instead, considering moderately-outflowing Comptonizing coronae and/or a more realistic high-spinning black hole population significantly relax the tension between the strength of the observed and modeled X-ray emission, while also predicting very low intrinsic scatter in the $L_X-L_{UV}$ relation. In particular, this latter scenario traces a known selection effect of flux-limited samples that preferentially select high-spinning, hence brighter, sources.
  
}

   \titlerunning{}
   \authorrunning{R. Arcodia et al.} 
   \maketitle

%
\defcitealias{Liu_zhu+2016:XMM-XXL}{L16}
\defcitealias{Shakura+Sunyaev1973:accretion}{SS73}
\defcitealias{Merloni2003:model}{M03}
\defcitealias{Lusso&Risaliti2017:toymodel}{LR17}
\section{Introduction}

The development of an in-depth understanding of accretion physics in active galactic nuclei (AGN) has tended to lag behind in comparison to other accreting objects (e.g., X-ray binaries, cataclysmic variables, and protoplanetary disks), for which many more observational constraints are available. While it seems that the standard thin-disk model \citep[][hereafter \citetalias{Shakura+Sunyaev1973:accretion}]{Shakura+Sunyaev1973:accretion} is not able to fully explain the plethora of accreting sources that we observe \citep[e.g.,][]{Koratkar+Blaes1999:status_accrdisks,Blaes2007:accr_discs_status,Antonucci2015:AGNpuzzle}, it is still unclear to what extent this simple but effective prescription has to be improved \citep{Kishimoto+2008:polarAD,Capellupo+2015:AGN_fit_ADs,Capellupo+2016:AD_fits_2}. 

Since the first AGN X-ray spectral surveys were performed \citep[e.g.][]{Elvis+1978:X_seyferts,Turner+Pounds1989:agn_survey_X}, the need for an additional spectral component to extend the cold-disk's $\lesssim\,$keV temperatures was evident. This so-called X-ray "corona" \citep[e.g.][]{Liang+Price1977:coronae,Galeev1979:coronae} is now almost universally considered as a hot ($\sim10^9\,$K), optically thin ($\tau\lesssim1$) plasma up-scattering the disk photons via thermal Comptonization \citep{Haardt+Maraschi1991:twophase1,Haardt+Maraschi1993:twophase2,Haardt+1994ApJ:patchy,Stern+1995:Xray_geom}, although an additional warm component is sometimes needed to fit the softest X-rays \citep[][and references therein]{Petrucci+2018:warm_corona,Kubota+Done2018:model_lx_luv}. The proximity of the corona to the central black hole was immediately suggested by its strong and fast variability \citep[e.g.][]{McHardy1989:variability} and by the reflection signatures \citep{Lightman+White1988:Xraybump,Pounds+1990:xray_reflect,Nandra1991:reflection,Williams+1992:ginga_spectra,Tanaka1995:relativistic_iron}, but in-depth information regarding its geometry and formation mechanism is still lacking. 

The geometry of the corona can be constrained via the observation of X-ray reverberation lags \citep{Fabian+2009:ironKlag,DeMarco+2013:softlags,Uttley2014:xrayreverb_review,Fabian2017:xray_reverb}, that seem to show a, possibly non-static, corona extending vertically and radially over the underlying disk for a few and a few tens of gravitational radii, respectively \citep[]{Wilkins+2016:modeling_Xreverb}. The compactness of the corona and the origin of the X-rays close to the black hole also appear to be confirmed by micro-lensing results \citep[e.g.][]{Mosquera+2013:lensedQSO_corona,Reis+2013:size_reverb_micro}. 

As far as the formation of the corona is concerned from the theoretical point of view, the most likely explanation for it is that it is magnetically-dominated with an efficient saturation of the magnetic field that is amplified via the magneto-rotational instability \citep[MRI,][]{Chandrasekhar1960:MRI_foreseen,B+H1991:MRI1,B+H1992:MRI4,H+B1991:MRI2,H+B1992:MRI3} and extending buoyantly upward (and downward) from the denser parts of the disk \citep{Galeev1979:coronae,Stella+Rosner1984:Binstabilities,DiMatteo1998:magn_reconnection,Merloni+Fabian2002:model_corona,Blackman+2009:coronae_largeB}. Magnetic reconnection can then keep the corona hot \citep[e.g.,][]{Liu+2002:B_corona,Uzdensky+2008:magnetized_corona_loops,Uzdensky+2016:magn_reconn,Beloborodov2017:magn_reconn,Werner+2019:reconn_IC,Ripperda+2019:MHDreconn}. This scenario seems to be supported by magneto-hydrodynamic (MHD) simulations \citep{Miller&Stone2000:corona_form,Uzdensky2013:simul_corona,Bai+Stone2013:accr_disc_MRI_corona,Jiang+2014:formation_coronae,Salvesen+2016:netpoloidalfield,Kadowaki+2018:magn_reconnection_corona}, although only qualitative comparisons with observations have been made so far \citep[however, see][]{Schnittman+2013:coronae_stellar_massBH}. Much effort has, nonetheless, been put into trying to shed light on the physics of the disk-corona system \citep[see][]{Blaes+2014:BHaccr_review} and this will continue with global 3D radiation-MHD simulations \citep[e.g.][]{Jiang+2017:globalSUPEREDD}, that are now approaching sub-Eddington flows as well \citep{Jiang+2019:subEdd_disks}. 

Observationally, the increase in quality and quantity of available AGN X-ray-to-UV data from large samples can provide insightful, and more easily approachable, diagnostics. The smoking gun of the disk-corona interplay in radiatively efficient AGN is given by the non linear correlation observed between the $2\,$keV and $2500\AA$ monochromatic luminosities \citep[e.g.,][and references therein]{Vignali+2003:alphaOX,Strateva+2005:alphaOX,Steffen+2006:alphaOX,Young2009:XMM_quasars,Lusso+2010:alphaOX,Lusso&Risaliti2016:LxLuvtight}, that persists throughout the common observed X-ray and optical-UV bands \citep{Jin+2012:optX_correlation}. Despite the possible differences arising from different sample selections and regression techniques, most observations point towards a $\log L_{X}-\log L_{UV}$ correlation with a slope $\approx0.6$, a dispersion that can be as small as $\sigma\approx0.2\,$dex \citep{Lusso&Risaliti2016:LxLuvtight,Chiaraluce+2018:dispandvariab_Lx_Luv}, and no apparent redshift dependency. Such a tight correlation paved the way for quasars to provide an alternative standard candle for cosmographic studies
\citep{Risaliti&Lusso2015:Hubble_diagram,Risaliti&Lusso2018:cosmo2}. The slope, being smaller than unity, indicates that from lowly to highly accreting AGN, the disk emission increases more than the corona emission \citep[e.g.,][]{Kelly+2008:AGNcorrelations} with crucial implications for the physics governing the coupled disk-corona system. However, a solid and conclusive theoretical explanation, for what is one of the most studied multi-wavelength observables in AGN, is still lacking.

The goal of this paper is indeed to test a self-consistently coupled disk-corona analytic model against the observed $L_{X}-L_{UV}$. Given the existing gap between simulations and observations, we argue that the use of simplified (but motivated) prescriptions still represents a powerful tool to explain observed disk-corona scaling relations, as it was done with the X-ray photon index (or the X-ray bolometric correction) correlation with the Eddington ratio \citep{Wang2004:hotdisccorona_constraints,Cao2009:coronamodel,Liu2009:disc-corona_investigated,You2012:model_disc_corona,Liu+2012:corona_model_highL,Liu+2016:structure_spec_corona_highL,Wang+2019:testing_stress}, or with the $\log L_{X}-\log L_{UV}$ itself (\citealp{Lusso&Risaliti2017:toymodel}, hereafter \citetalias{Lusso&Risaliti2017:toymodel}; \citealp{Kubota+Done2018:model_lx_luv}). We here rely uniquely on the $\log L_{X}-\log L_{UV}$ relation, since monochromatic $L_{X}$ and $L_{UV}$ values can be directly obtained from spectral fits. Forward modeling monochromatic luminosities circumvents difficulties and issues typical of model comparisons with accretion rate, Eddington ratio or bolometric luminosity estimates \citep[e.g.,][]{Richards+2006:qso_sed,Davis&Laor2011:efficiency_Mdot,Slone&Netzer2012:winds_caveat_Lbolandco,Krawczyk+2013:BC_caveats,Capellupo+2015:AGN_fit_ADs,Capellupo+2016:AD_fits_2,KilerciEcer2018:BC_caveats}. 

We describe our disk-corona model in Section~\ref{sec:model} (and Appendix~\ref{sec:app_model}) and we briefly show its qualitative predictions in Section~\ref{sec:prediction_lxluv}. Then, we outline the observational test that we put forward to thoroughly understand the disk-corona interplay in Section~\ref{sec:observational_test} and we show the results in Section~\ref{sec:results}. Throughout this work, we quote median values with 16th and 84th percentiles unless otherwise stated.

\section{The disk-corona model}
\label{sec:model}

The disk-corona model adopted in this work is largely based on the prescriptions put forward by \citet[][hereafter \citetalias{Merloni2003:model}; see also \citealp{Merloni+Fabian2002:model_corona,Merloni&Fabian2003:coronaGR}]{Merloni2003:model}, in which the standard conservation equations of a geometrically-thin and optically-thick accretion disk \citep[\citetalias{Shakura+Sunyaev1973:accretion};][]{Pringle1981:accr_disc} are self-consistently coupled with the X-ray corona, indicated as the fraction $f$ \citep[e.g.,][]{Haardt+Maraschi1991:twophase1,Svensson&Zdziarski94:corona_f} of accretion power (per unit area, $Q_+$) that is dissipated away from the cold disk \citep[e.g.,][]{Stella+Rosner1984:Binstabilities,DiMatteo1998:magn_reconnection}:
\begin{equation}
\label{eq:eqf}
f=\frac{Q_{cor}}{Q_+}
\end{equation} with $Q_{cor}=v_DP_{mag}$ and $Q_{+}=\frac{3}{2}c_s\tau_{r\phi}$,  where $v_D$ is the vertical drift velocity (taken proportional to the Alfvén speed via an order-unity constant $b$), $P_{mag}=B^2/8\pi$ is the magnetic pressure, $c_s$ is the sound speed and $\tau_{r\phi}$ is the vertically-averaged stress tensor. 

The stress tensor can be assumed to be dominated by Maxwell stresses \citep[e.g.,][]{Hawley+1995:Bstresses,Sano+2004:MHD_accrdisc,Minoshima+2015:MRI_gaspress}, from which we can write $\tau_{r\phi}=k_0 P_{mag}$, with $k_0$ being a constant of order unity \citep{Hawley+1995:Bstresses}. To build a self-consistent solution to the accretion problem, we need to relate the stress tensor (via the magnetic pressure) with local quantities that standard analytic models are familiar with. As a matter of fact, the $\alpha-$prescription is not the only educated guess that is adopted to dodge our ignorance of the physical mechanism producing the disk viscosity. Within the same theoretical framework, fundamental modifications to the viscosity law can be introduced depending on whether the viscous stress is assumed to scale proportionally with the total ($P_{tot}$, gas plus radiation) pressure \citepalias{Shakura+Sunyaev1973:accretion}, with the gas pressure alone \citep{Lightman+Eardley1974:instability,Sakimoto+Coroniti1981:viscosity,Meyer+M-H1982:verticalstructure,Stella+Rosner1984:Binstabilities} or with the geometric mean of the two \citep[\citetalias{Merloni2003:model};][]{Ichimaru1977:accretion,Taam+Lin1984:viscous_accretion,Burm1985:scaling_stress}. It was soon discovered that the first prescription leads to thermally and viscously unstable disks in the radiation-pressure dominated regions, with the first instability acting on shorter timescales \citep{Lightman+Eardley1974:instability,SS1976:instab,Pringle1976:therm_instab}. This encouraged many authors \citep{Hoshi1985:generalized_viscosity,Szuszkiewicz1990:viscosity_laws,Merloni+Nayakshin2006:viscosity_laws,Gredzielski+2017:mod_viscosity} to generalize the viscosity law. Recent simulations (albeit of gas-pressure dominated disks only) indeed seem to show a power-law stress-pressure relation \citep{Sano+2004:MHD_accrdisc,Minoshima+2015:MRI_gaspress,Ross+2016:stress_vs_pressure,Shadmehri+2018MNRAS:stress_vs_press}, with an index varying from zero to one according to the different assumptions. 

Here, we address this issue generalising the model reported in \citetalias{Merloni2003:model} with:
\begin{equation}
\label{eq:viscosity_law}
P_{mag}=\alpha_0P_{gas}^{\mu}P_{tot}^{1-\mu}
\end{equation}
where $\alpha_0$ is a constant, generally not equal to $\alpha_{SS73}=P_{mag}/P_{tot}$. This behavior is physically motivated by the MRI prescriptions, as its growth rate was shown to depend on the $P_{rad}$-to-$P_{gas}$ ratio \citep{Blaes&Socrates2001:Prad_MRI,Turner+2002:Prad_MRI_simul} influencing the level of the magnetic field saturation. Equations~\ref{eq:eqf} and~\ref{eq:viscosity_law} provide the closure equation of the disk-corona system:

\begin{equation}
\label{eq:eqf_closure}
f=\sqrt{\frac{2\alpha_0}{k_1^2}}\left(1+\frac{P_{rad}}{P_{gas}}\right)^{-\mu/2}\end{equation}where $k_1=3k_0/2b$ gathers the model's uncertainties in an order unity factor \citepalias{Merloni2003:model}. Its exact value only affects $f$ at its maximum $\Big(f_{max}=\sqrt{2\alpha_0/k_1^2}\Big)$ and not the nature of what is described throughout this paper.

The model is then completed with the equation of state:\begin{equation}
\label{eq:eqstate}
P_{tot}=P_{gas}+P_{rad}= \frac{2\rho k_BT}{m_p}+\frac{aT^4}{3}
\end{equation} and with a density- and temperature-dependent opacity $\kappa=\kappa\,(\rho,T)$. We compute the opacity value self-consistently with the density and temperature at each radius with an iterative process, using as reference stellar opacity tables (at solar metallicity) from the Opacity Project \citep{Seaton+1994:OP_project,Seaton+1995:OP_2}. This is important since the density and temperature regimes relevant for AGN disks imply opacities that can be significantly different from the electron scattering value \citep[e.g., see][]{Jiang+2016:iron_bump_AGN,Czerny+2016ApJ:BLR_newOP,Grzedzielski+2017:opacity}. 

Further, we assume a downward component of the X-ray emission ($\eta$) and a disk albedo ($a_{disk}$), which modify the disk equations from the usual ($1-f$) factor \citep[\citetalias{Merloni2003:model};][]{Svensson&Zdziarski94:corona_f} to:\begin{equation}
\label{eq:eqf_tilde}
1-\tilde{f}=1-f\left[1-\eta\left(1-a_{disk}\right)\right]
\end{equation}We here adopt $\eta=0.55$ and $a_{disk}=0.1$, respectively \citep[e.g.,][]{Haardt+Maraschi1993:twophase2}. These are typical values for anisotropic Comptonization in plane-parallel geometry, although more generally the product $\eta\left(1-a_{disk}\right)$ can be a function of the photon index $\Gamma$ \citep{Beloborodov+1999:HE_accr,Malzac+2001:dyn_coronae} and of the disk's vertical structure.

For simplicity, we adopt dimensionless units for the black hole mass, the accretion rate, the radial distance and the vertical scale-height:
\begin{equation}
	\begin{cases}
		m=M/M_{\astrosun} \\
		\dot{m}=\frac{\dot{M}}{L_{edd}/\epsilon_0c^2}=m^{-1} \frac{\dot{M}\epsilon_0c^2}{1.3\times10^{38}} \\
		r=R/R_s=m^{-1}Rc^2/2GM_{\astrosun} \\
		h=H/R_s=m^{-1}Hc^2/2GM_{\astrosun} 
	\end{cases}
\end{equation}

The equations for $h$, mid-plane $\rho$ (g\,cm$^{-3}$), $P$ (dyn\,cm$^{-2}$) and $T$ (K), with the closure equation for $f$, are reported in Appendix~\ref{sec:app_model} in Newtonian approximation \citep[however, see][for a relativistic derivation of the $\mu=0.5$ case]{Merloni&Fabian2003:coronaGR}, along with the related radial profiles (Fig.~\ref{fig:other_profiles}). We note that a constant efficiency of $\epsilon_0=0.057$, typical of non-rotating black holes, and a no-torque inner boundary condition ($J(r)=1-\sqrt{r_0/r}$, with $r_0=3$ and $r_{out}=2000$) are initially adopted. 

Once $m$, $\dot{m}$, $\alpha_0$, $\mu$ and $f_{max}$ are fixed, one can numerically solve the closure equations for $f$ at each radius (see the last rows of Eq.~\ref{eq:system_rad_mu} and~\ref{eq:system_gas}, respectively). The left-hand side is equal to $P_{rad}/P_{gas}$ and we can infer the correct regime and compute the main physical quantities at the mid-plane ($\rho$, $P$, $T$, $\kappa$). Then, the effective temperature at the surface is computed:\begin{equation}
\label{eq:Teff}
T_{eff}(r)\propto\frac{T(r)}{\tau(r)^{1/4}}
\end{equation}where we take $\tau(r)=h(r)\,\rho(r)\,\kappa(r)$. Monochromatic optical-UV luminosities can be then easily computed in the multi-color blackbody approximation:\begin{equation}
\label{eq:Luv_BB}
L_{\nu}(r)=2\pi r\Delta r\,\pi\,B_{\nu}(T_{eff})
\end{equation}where $\pi\,B_{\nu}(T_{eff})$ is the black-body flux at the frequency $\nu$ and temperature $T_{eff}(r)$. 

\begin{figure}[tb]
	\centering
	\includegraphics[width=0.95\columnwidth]{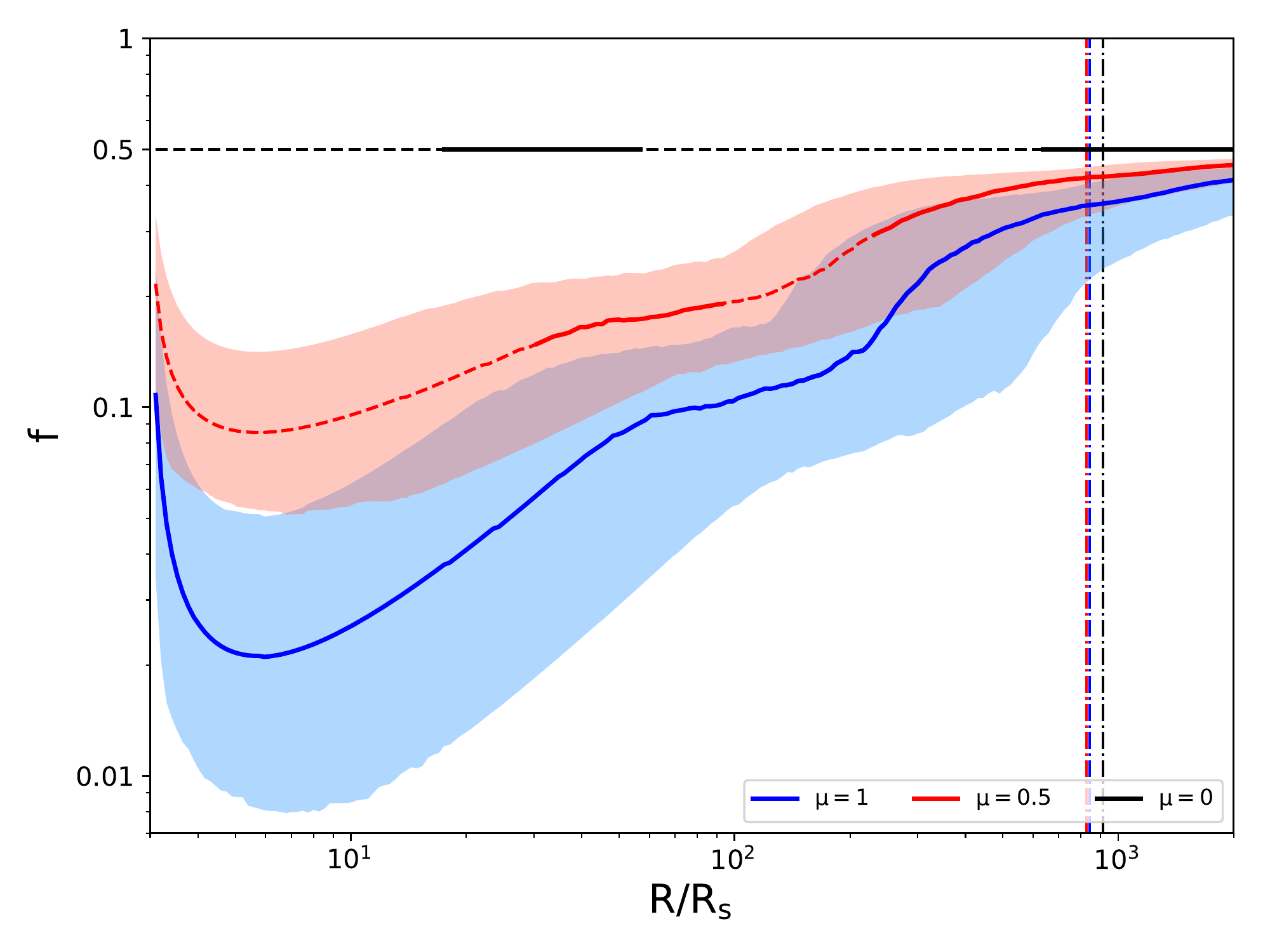}
	\includegraphics[width=0.95\columnwidth]{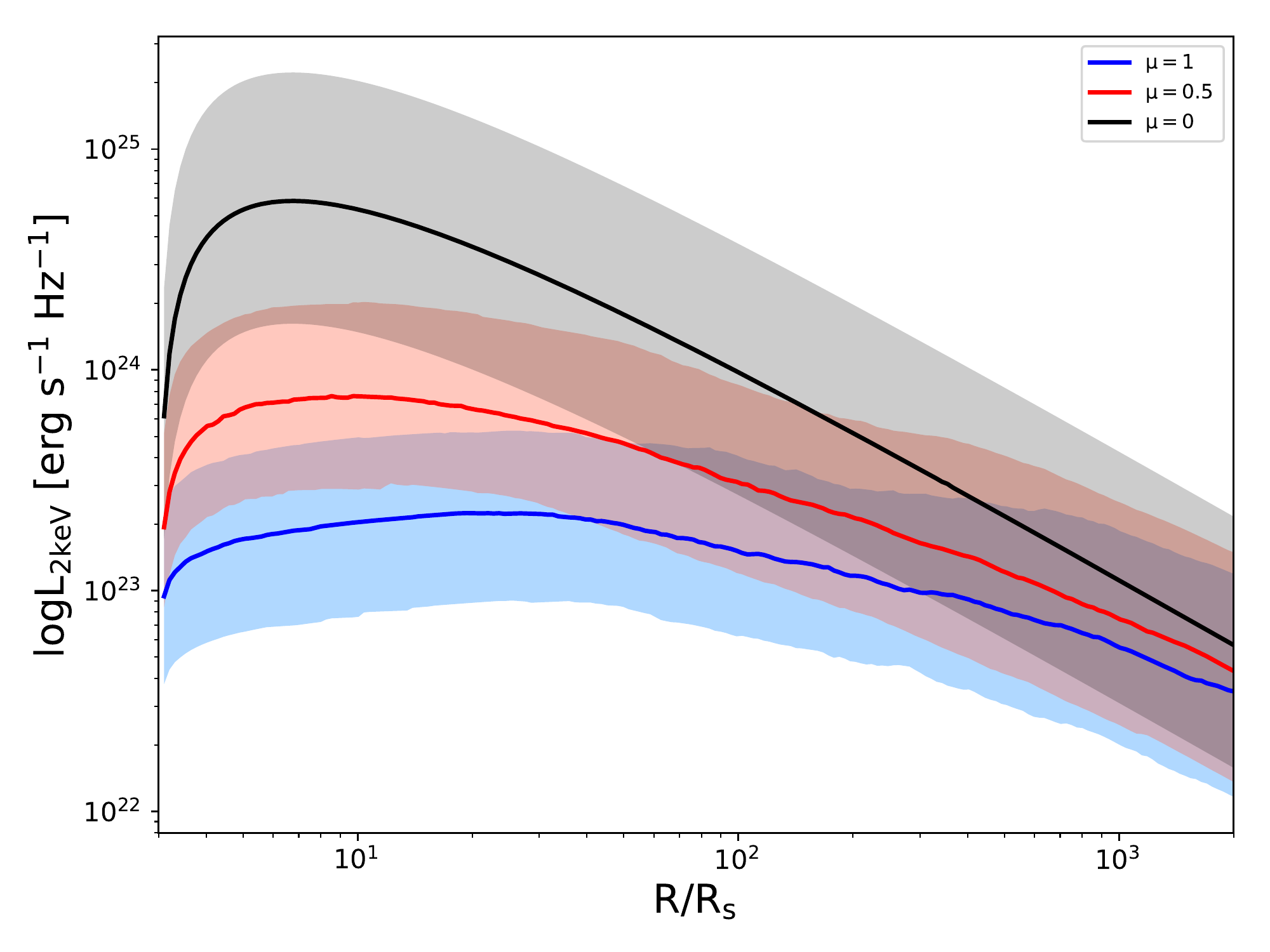}	\caption{Radial profiles for the fraction $f$ of power dissipated in the corona (top panel) and $L_{2keV}$ (bottom panel), obtained with fixed $\alpha_0=0.02$ and $f_{max}=0.5$. Colors are coded according to the choice of the viscosity law: stress proportional to $P_{tot}$ ($P_{gas}+P_{rad}$, $\mu=0$, black), to $P_{gas}$ ($\mu=1$, blue), or the geometric mean of the two ($\mu=0.5$, red). The continuous solid, or solid-dashed, lines represent the median profiles, with the related shaded areas showing the 16th and 84th percentiles, the scatter due to the range of sources (i.e. $m$ and $\dot{m}$) modeled. $L_{2keV}$ is proportional to the product of $f$ and $Q_{+}$ (the accretion power per unit area). As $Q_{+}$ has very similar profiles across all models, those systems for which $f$ is smaller produce weaker coronae in the central part. In the top panel, a solid line for the median $f$-profile represents (thermally) stable regions of the median test source, whereas a dashed line highlights the instability regions. The vertical dot-dashed lines show instead where the median transition radius, from $P_{rad}$- to $P_{gas}$-dominated regions, lies. Refer to Section~\ref{sec:radial_profiles} for details.}
	\label{fig:f_Lx_profiles}
\end{figure}

In this framework, the energy per unit area dissipated in the corona at each radius is $Q_{cor}(r)=f(r)Q_{+}(r)$, although only a fraction $(1-\eta)$ will contribute to what is observed as X-ray emission:\begin{equation}
\label{eq:LX_tot}
L_{X,tot}(r)=2\pi r\Delta r\,(1-\eta)\,f(r)Q_{+}(r)
\end{equation}Here, we did not include the component reflected by the disk (given by the fraction $f\,\eta\, a_{disk}$), so that we could easily extrapolate at each radius a monochromatic value, for instance $L_{2keV}$, assuming a simple power-law spectrum within $\nu_i=0.1$ and $\nu_f=100\,$keV:\begin{equation*}L_{X,tot}=K\int_{\nu_i}^{\nu_f}\nu^{-(\Gamma-1)}\,d\nu
\end{equation*}\begin{equation}
\label{eq:L2kev_extrapolation}
L_{2keV}=K\,{\nu_{2\,keV}}^{-(\Gamma-1)}=L_{X,tot}\,(2-\Gamma)\,\frac{{\nu_{2\,keV}}^{1-\Gamma}}{{\nu_f}^{2-\Gamma}-{\nu_i}^{2-\Gamma}}
\end{equation}

The model relies on the assumption that a plane-parallel geometry holds for bright radiatively-efficient sources, lying in a sweet spot of accretion rate ($\dot{m}$ approximately from a few percent to Eddington). Hence, the accretion disk extends down to the innermost stable circular orbit (ISCO) and no advection is included. A scripted version of the model outlined in this Section will be made publicly available online\footnote{\href{https://github.com/rarcodia/DiskCoronasim}{https://github.com/rarcodia/DiskCoronasim}}.

\subsection{Radial profiles for $f$ and $L_{2keV}$}
\label{sec:radial_profiles}

In Fig.~\ref{fig:f_Lx_profiles} we show as an example radial profiles of $f$ and $L_{2keV}$, obtained by solving Eq.~\ref{eq:system_rad_mu},~\ref{eq:system_gas}, ~\ref{eq:LX_tot} and~\ref{eq:L2kev_extrapolation}. For simplicity, we fixed $\alpha_0=0.02$ and $f_{max}=0.5$ and used the three values of $\mu$ corresponding to the most-used viscosity laws, namely $\mu=0$, 0.5 and 1, for $P_{mag}$ proportional to $P_{tot}$, $\sqrt{P_{gas}P_{tot}}$ and $P_{gas}$, respectively. Other values of $\mu$ would support the same picture with analogous intermediate profiles. A range of typical $m$, $\dot{m}$ and X-ray spectral slopes was chosen, following the distribution of objects observed in the survey field adopted for the observational test (see Section~\ref{sec:observational_test} below), namely with median values (and related 16th and 84th percentiles) of $\log m=8.7_{-0.5}^{+0.4}$, $\dot{m}=0.2_{-0.1}^{+0.5}$ and $\Gamma=2.1\pm0.1$.

The solid (or solid-dashed) lines represent the median profiles, with the corresponding shaded areas showing the 16th and 84th percentiles of the distribution. The top panel of Fig.~\ref{fig:f_Lx_profiles} shows how the standard $\mu=0$ law (e.g., \citetalias{Shakura+Sunyaev1973:accretion}) results in $f=f_{max}$ at all radii (e.g., \citealp{Svensson&Zdziarski94:corona_f}), whereas alternative viscosity laws (e.g., $\mu=0.5$ and $\mu=1$) show non-constant radial profiles for $f$: in the latter cases, the fraction of power dissipated in the corona is smaller in the regions strongly dominated by $P_{rad}$.  As it was shown in \citetalias{Merloni2003:model} in particular for the $\mu=0.5$ scaling, the higher suppression of the growth rate in $P_{rad}$-dominated regions of the disk leads to such damped $f$-profiles. This directly influences the strength of the corona emission, as $L_{2keV}$ is proportional (through $L_{X,tot}$) to the product of $f$ and $Q_{+}$: $Q_{+}$ peaks at small radii in a very similar way across all models, therefore the ones with deeper $f$-profiles show flatter $L_{2keV}$ radial profiles and, hence, weaker coronae (bottom panel of Fig.~\ref{fig:f_Lx_profiles}). The exact shape of $f(r)$ also affects the strength of the disk emission since the two are self-consistently coupled (see Eq.~\ref{eq:eqf_tilde}).

We can also define the mean value of each $f(r)$ profile (i.e. for each combination of $m$, $\dot{m}$ and $\Gamma$):\begin{equation}
\label{eq:f_mean}
\text{<}f\text{>}_i=\frac{\int f_i(r)\,Q_{+,i}(r)\,2\pi r\,dr}{\int Q_{+,i}(r)\,2\pi r\,dr}
\end{equation}that is also a function of $\mu$, $\alpha_0$ and $f_{max}$. Then, the mean value can be computed for the median $f$ profiles in the examples in the top panel of Fig.~\ref{fig:f_Lx_profiles}: $\text{<}f\text{>}_{median}=0.5$, 0.13 or 0.05, for $\mu=0$, 0.5 and 1, respectively. Of course, within such a model the exact value of $\text{<}f\text{>}_{median}$ depends on its normalization $f_{max}$, that is a free parameter in the model only bound to be $<1$. Nonetheless, simply from looking at $\text{<}f\text{>}_{median}$ as a function of $\mu$ (and from Fig.~\ref{fig:f_Lx_profiles}) we can see how, for the same set of inputs (e.g., $m$, $\dot{m}$), the different accretion prescriptions relate to the output corona luminosities: in a nutshell, going from $\mu=0$ to $\mu=1$ produces lower $\text{<}f\text{>}_{median}$, thus weaker coronae. 

Changing $\mu$ also affects the logarithmic scatter in the radial profiles, from being absent in $\mu=0$ to increase with $\mu$ for $\mu\neq0$ (see Fig.~\ref{fig:f_Lx_profiles}). The spread on a given $f(r)$ is due to the scatter in $m$, $\dot{m}$ and $\Gamma$, where the major role is played by the accretion rate (e.g., see Fig. 1 in \citetalias{Merloni2003:model}). Crucially, $\text{<}f\text{>}$ decreases with increasing $\dot{m}$ for all $\mu\neq0$ models. This points in the same direction as the evidence of an X-ray bolometric correction (that is proportional to the inverse of $f$) increasing with the accretion rate \citep[e.g.][]{Wang2004:hotdisccorona_constraints,Vasudevan+Fabian2007:kbol_1,Vasudevan+Fabian2009:kbol_2,Lusso+2010:alphaOX,Young+2010:alpha_ox_kbol}. This relation between $\text{<}f\text{>}$ and $\dot{m}$ has also crucial implications for what our models predictions on the physical mechanisms behind the observed $L_X-L_{UV}$ (see Section~\ref{sec:prediction_lxluv}).



\subsection{Local thermal stability}

Before proceeding to a detailed observational test of the model, we briefly discuss here the stability issue for the various adopted viscosity laws. \citet{Jiang+2016:iron_bump_AGN} showed that the presence of the iron bump in the opacity at $\sim2\times10^5\,$K stabilizes the flow in the disk regions around that temperature, where the cooling term has a different dependency and thermal runaway is avoided \citep{Grzedzielski+2017:opacity}. In the top panel of Fig.~\ref{fig:f_Lx_profiles}, a solid median line for the $f$-profile represents (thermally) stable regions of the median test source, whereas a dashed line highlights the instability regions. The vertical dot-dashed lines show instead where the median transition radius, from $P_{rad}$- to $P_{gas}$-dominated regions, lies. This highlights that, for the median test source, the stability region extends also well within $P_{rad}$-dominated regions of the disk, confirming previous results \citep{Jiang+2016:iron_bump_AGN,Grzedzielski+2017:opacity}. More quantitatively, we computed the thermal stability balance \citep[e.g.][]{Pringle1976:therm_instab} for each test source ($m$, $\dot{m}$) at all radii with varying viscosity laws. The new stability regions in the inner $P_{rad}$-dominated portions of $\mu=0$ and $\mu=0.5$ disks are ubiquitous, but they appear at different radii according to where the disk reaches the temperatures around the iron bump in $\kappa$ (see also Fig.~\ref{fig:other_profiles}). The $\mu=1$ case, as it is well known \citep[e.g.][]{Lightman+Eardley1974:instability}, is stable throughout. 

\section{Predictions of the model on the $L_X-L_{UV}$ relation}
\label{sec:prediction_lxluv}

\begin{table*}[tb]
	\normalsize
	\caption{Summary of recurring model parameters.}
	\label{tab:model_param}
	\centering
	\begin{tabular}{C{0.07\textwidth} C{0.4\textwidth} C{0.25\textwidth}}%
		\toprule
		\multicolumn{1}{c}{Parameter} &
		\multicolumn{1}{c}{Definition} &		
		\multicolumn{1}{c}{Comments} \\
		\midrule
		& Regulates the scaling between the magnetic stress & $\mu=0\,\,$ $\rightarrow$ $\,\,P_{mag}\propto P_{tot}$ \\
		$\mu$ &  and the thermal pressure: & $\mu=0.5\,\,$ $\rightarrow$ $\,\,P_{mag}\propto \sqrt{P_{gas} P_{tot}}$ \\ 
		& $\tau_{r\phi}\propto P_{mag}=\alpha_0P_{gas}^{\mu}P_{tot}^{1-\mu}$ & $\mu=1\,\,$ $\rightarrow$ $\,\,P_{mag}\propto P_{gas}$ \\ 
		\midrule
		$\alpha_{0}$ & Proportionality constant of the viscosity law (see above) & Small influence on $L_X-L_{UV}$\\
		\midrule
		$f(r)$ & Fraction of accretion power dissipated in the corona & $f(r)$ for $\mu\neq0$ \\		
		\midrule
		$f_{max}$ & Maximum value of $f(r)$ & Impacts the normalization of the $L_X-L_{UV}$\\
		\midrule
		$\text{<}f\text{>}$ & Mean value of a $f(r)$ profile & Real fraction of bolometric power emitted by the corona \\
		\midrule
		$\eta$ & Fraction of $f$ emitted downward back to the disk & Exact value impacts the normalization of the $L_X-L_{UV}$ \\
		\bottomrule	
	\end{tabular}
\end{table*}

In this section we aim to test our disk-corona model (presented in Section~\ref{sec:model}) against the observed $L_{X}-L_{UV}$, a robust observable linked to the disk-corona physics. Before performing a more quantitative observational test (Section~\ref{sec:observational_test}), we here outline the predictions of our model concerning the disk-corona energetics and the expected impact of our accretion prescription on the $L_{X}-L_{UV}$.

\begin{figure}[tb]
	\centering
	\includegraphics[width=0.99\columnwidth]{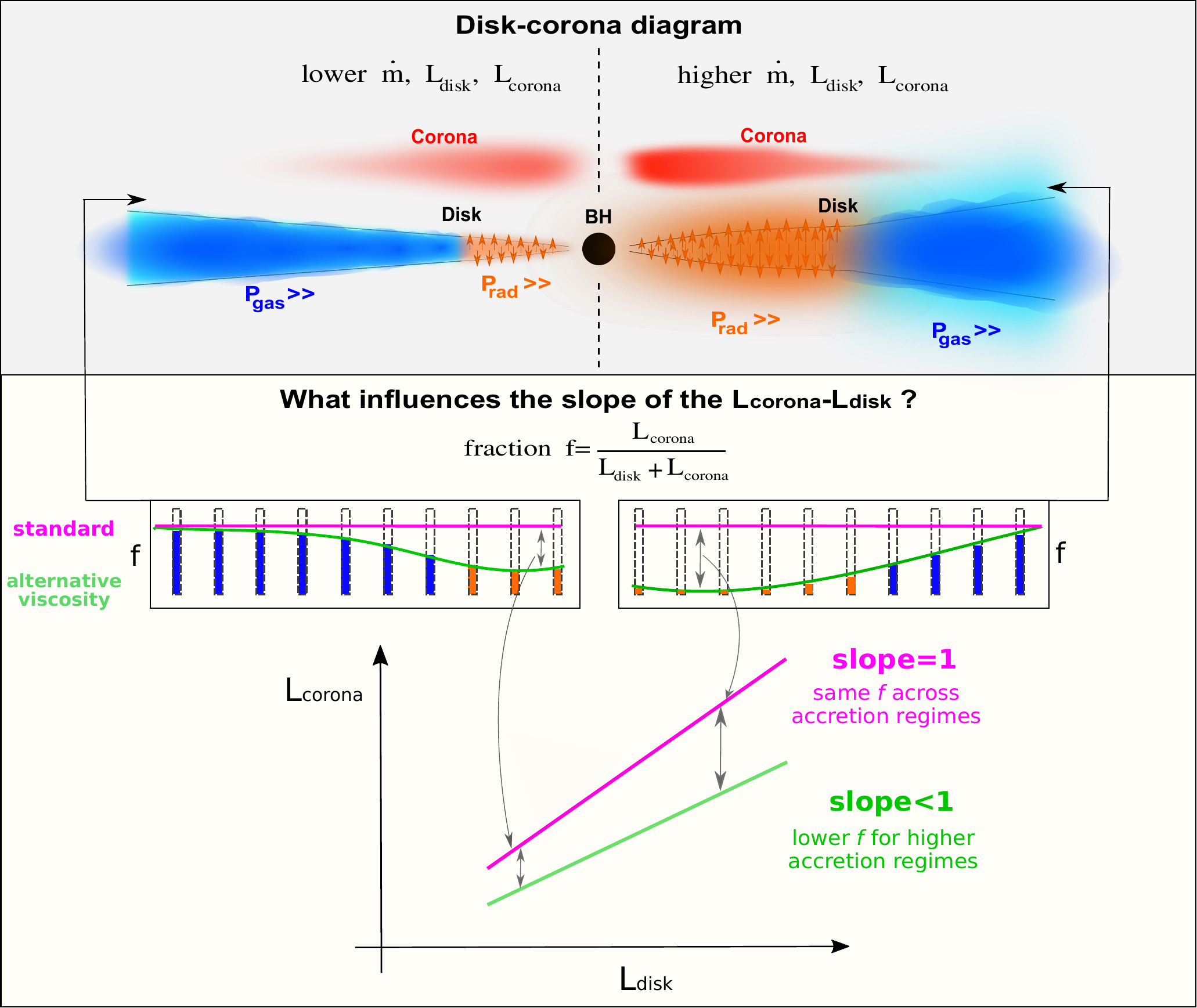}	
	\caption{Schematic illustration of our model and how it relates to the observed $L_X-L_{UV}$ (see Section~\ref{sec:prediction_lxluv} for an interpretative guide).}
	\label{fig:cartoon}
\end{figure}

The schematic illustration in Fig.~\ref{fig:cartoon} summarizes the qualitative take-home messages of this work. The observed $L_{X}-L_{UV}$ states that going from a lower to a higher accretion regime, the luminosity of the corona increases less than the disk luminosity, resulting in a slope smaller than one in the log-space. In our model for the disk-corona system, the luminosity outputs are directly modified by the viscosity prescription in the flow, determined by the parameter $\mu$, and by the fraction of accretion power going into the corona, $f$ (see Table~\ref{tab:model_param} for a summary on the model parameters). Among all scenarios spanned by these two main unknowns, the qualitative behavior of the accretion disk-corona system, along its radial extent, is similar: higher accretion states have a more powerful disks and coronae, but wider $P_{rad}$-dominated inner region and, only for modified viscosity prescriptions (i.e. $\mu\neq0$), lower relative contribution of the corona to the total luminosity (see the upper diagram in Fig.~\ref{fig:cartoon}).

Thus, our model can provide a simple explanation for the observed slope of the $L_X$-$L_{UV}$ relation, bridging in a simple but effective way the gap between the observed X-to-UV energetics and some aspects of MRI simulations. Changing $\mu$ not only affects the disk thermodynamics, but also changes the amount of power carried away by the corona (see Fig.~\ref{fig:f_Lx_profiles}). A constant radial profile for $f$ (e.g., \citealp{Svensson&Zdziarski94:corona_f}; here $\mu=0$) would naturally result in a $L_{X}-L_{UV}$ close to a one-to-one relation. On the contrary, the alternative viscosity prescriptions, that we identify with $\mu\neq0$, inherently result in a different disk-corona energetic coupling for varying accretion rates: in particular, higher $\dot{m}$ yield more damped $f-$profiles (see also \citetalias{Merloni2003:model}). In this scenario, the outcome would  be a slope of the $L_{X}-L_{UV}$ that is smaller than one (see the lower diagram in Fig.~\ref{fig:cartoon}). 

It is worth stressing that it is only the relative fraction $f$ that is more suppressed in the inner regions of systems with a modified viscosity and not the X-ray emission per se. Regardless the underlying assumption of a plane-parallel geometry for the disk-corona system (see Eq.~\ref{eq:eqf_tilde}), the X-ray emission peaks in the innermost radii (e.g., see the $\mu=0.5$ case in Fig.~\ref{fig:f_Lx_profiles}). This will be addressed in details in Section~\ref{sec:results_norm}.

\begin{figure}[tb]
	\centering
	\includegraphics[width=0.9\columnwidth]{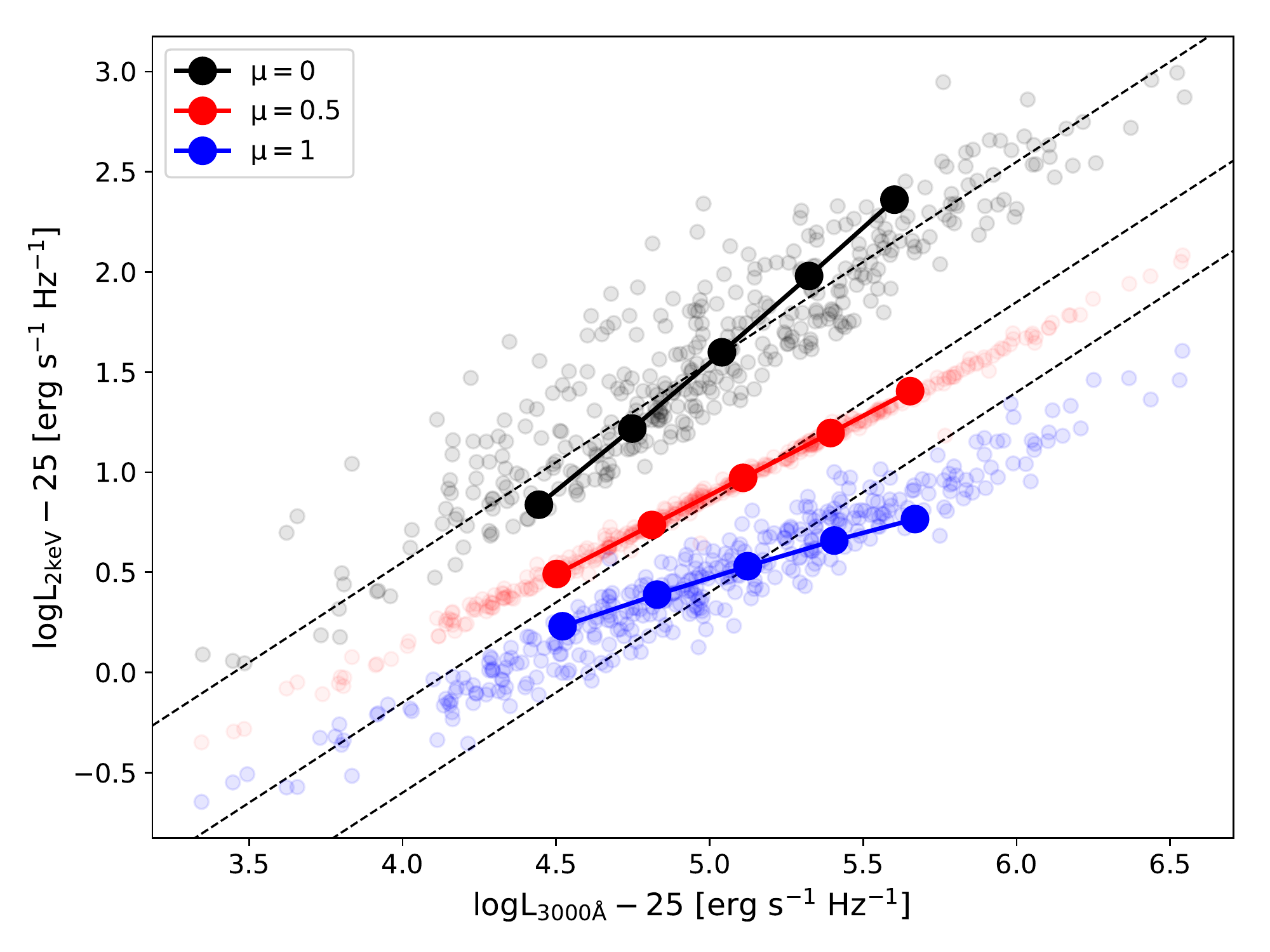}	
	\caption{Mock $L_{X}-L_{UV}$ with fixed $f_{max}=0.5$ and different $\mu$ color coded, as in Fig.~\ref{fig:f_Lx_profiles}. The connected solid points show the trend of a single typical mass ($\log m=8.7$) with $\dot{m}=0.03$, 0.07, 0.17, 0.42, 1 ( increasing from left to right in $L_{UV}$). The dashed lines indicate a slope of one. The distributions of transparent points show the mock $L_{X}-L_{UV}$ for a range of $\log m=8.7_{-0.5}^{+0.4}$, $\dot{m}=0.2_{-0.1}^{+0.5}$ and $\Gamma=2.1\pm0.1$, that follows the typically observed objects (see Section~\ref{sec:observational_test}).}
	\label{fig:single_mass_lx_luv}
\end{figure}

Fig.~\ref{fig:single_mass_lx_luv} shows an example of a mock $L_{X}-L_{UV}$ from the model realizations, with $f_{max}=0.5$ and different values of $\mu$ as in Fig.~\ref{fig:f_Lx_profiles}, with the same color coding. For the connected solid points, a single typical mass ($\log m=8.7$) is adopted, with increasing $\dot{m}=0.03$, 0.07, 0.17, 0.42, 1 (from left to right in $L_{UV}$). The relation is linear for a given mass, with the dashed lines indicating a slope of one to guide the eye. As qualitatively shown in the illustrative Fig.~\ref{fig:cartoon}, models where $f$ is constant in radius and in accretion state (black points) yield a slope close to one (even higher for the single mass); instead, alternative viscosity prescriptions (red and blue), that change the disk-corona energetic interplay via $f(r)$, show a flatter slope. The underlying transparent points show the mock $L_{X}-L_{UV}$ for a distribution of $\log m=8.7_{-0.5}^{+0.4}$, $\dot{m}=0.2_{-0.1}^{+0.5}$ and $\Gamma=2.1\pm0.1$, that follows the typically observed objects (see Section~\ref{sec:observational_test}).

\section{Observational test: modeling the $L_{X}-L_{UV}$}
\label{sec:observational_test}

In the previous section we qualitatively outlined what are the physical mechanisms identified as the origin of the observed slope smaller than one. A more quantitative test is needed to thoroughly investigate all aspects of the observed $L_{X}-L_{UV}$, including its normalization and intrinsic scatter. The exact value of the slope given by the models not only depends on the unknowns $\mu$, $f_{max}$ and $\alpha_{0}$, but also on the details of the distributions of $m$, $\dot{m}$, $\Gamma$ that are adopted for the calculations. Nonetheless, not all combinations of these three parameters are observed, because they do not exist in nature or we are biased against their detection (for example, the black masses of AGN follow a given mass distribution, some some mass ranges are more probable than others in any observed sample). That is why we select a reference sample of radiatively-efficient broad-line AGN (Section~\ref{sec:sample}) and model the most likely values for $m$, $\dot{m}$, $\Gamma$ for each source individually, based on the available data. 


\subsection{The reference sample of broad-line AGN}
\label{sec:sample}

We built our reference sample starting from the 1787 AGN within the XMM-XXL north survey \citep{Pierre+2016:XMM-XXL} identified as broad-line AGN (BLAGN) by the Baryon Oscillation Spectroscopic Survey (BOSS) follow-up \citep[][hereafter \citetalias{Liu_zhu+2016:XMM-XXL}]{Menzel+2016:XXM-XXL_boss,Liu_zhu+2016:XMM-XXL}. The X-ray spectral analysis on these sources was performed in \citetalias{Liu_zhu+2016:XMM-XXL} with the Bayesian X-ray Analysis software \citep[BXA,][]{Bucnher+2014:BXA_agnrefmodel}, providing $N_H$ values, photon indexes ($\Gamma$) and the rest-frame $2-10\,$keV intrinsic luminosities ($L_{2-10\,keV}$). Furthermore, single-epoch virial black hole masses \citep[$M_{BH}$, e.g.][]{Shen+2008:virMBH_biases} and continuum luminosities (at 1350, 1700, 3000 and $5100\,\AA$) were obtained on the BOSS spectroscopy with a fitting pipeline \citep[for details, we refer to \citetalias{Liu_zhu+2016:XMM-XXL};][]{Shen&Liu2012:fit_QSO_shengroup}. Luminosities were computed in \citetalias{Liu_zhu+2016:XMM-XXL} assuming $H_0=70\,$km s$^{-1}$ Mpc$^{-1}$, $\Omega_m=0.27$ and $\Omega_{\Lambda}=0.73$\footnote{We will refer to other data throughout the paper and possible discrepancies in luminosities due to different cosmological parameters may occur. Nonetheless, we verified that the biggest difference in luminosity values ($\sim 0.01~$dex) is obtained assuming a Planck with respect to a WMAP release, while using different releases of the same instrument will have a negligible impact ($\lesssim 0.005~$dex).}.

Then, we applied some cleaning criteria to avoid, as much as possible, imprecise estimates for the intrinsic (accretion-powered) $L_{X}$ and $L_{UV}$ and to remain consistent with what is computed by the model. Firstly, among the monochromatic luminosity values available in the optical-UV from \citetalias{Liu_zhu+2016:XMM-XXL} we adopted $L_{3000\AA}$, obviously inducing a redshift cut in the sample (see Fig.~\ref{fig:XMM-XXL_sensitivity} in Appendix~\ref{sec:Lx_Luv_XXL}). Model wise there is no difference in computing $L_{3000\AA}$ or the more standard $L_{2500\AA}$, and \citet{Jin+2012:optX_correlation} showed compatible correlations between the X-ray luminosity and each wavelength of the optical spectrum, although their coverage starts from $3700\AA$. We verified a posteriori that this choice does not affect significantly the slope of the $L_{X}-L_{UV}$ or the conclusions of our work.

Secondly, despite being defined as BLAGN, \citet{Liu_teng+2018:Xobs_type1AGN} found that a fraction of these sources shows continuum reddening probably due to intervening dust along the line of sight, not accounted for by our model. \citet{Liu_teng+2018:Xobs_type1AGN} defined a slope parameter $\alpha'$ for the optical-UV continuum, to discern between the reddened sources and the bulk of blue BLAGN at each redshift. The contamination from extinction at $L_{3000\AA}$ was minimized by conservatively selecting sources with $\alpha'<-0.5$ \citep[see][their Fig. 2]{Liu_teng+2018:Xobs_type1AGN}.

Moreover, the XMM-XXL survey has a typical exposure time of $\sim10\,$ks per pointing (\citetalias{Liu_zhu+2016:XMM-XXL}; \citealp{Pierre+2016:XMM-XXL}). Here, the analysis was restricted to sources with at least 10 counts in the EPIC-pn \citep{Struder:pn} and EPIC-MOS \citep{Turner:MOS} cameras on board XMM-Newton \citep{Jansen:XMM}, to exclude sources with extremely low-quality X-ray spectra.

Then, to exclude data contaminated by X-ray absorption, not accounted for in our modeling, we conservatively selected only sources in which the 84th percentile of the $N_H$ posterior distribution was smaller than $10^{21.5}\,$cm$^{-2}$, a value typically adopted to distinguish X-ray obscured and un-obscured sources (\citealp{Merloni+2014:obscuredAGN}; see also \citealp{DellaCeca+2008:AGNprop}).

Finally, we take $L_{2keV}$ as reference for the corona emission in the $L_{X}-L_{UV}$.	In the model, we computed mock $L_{X,tot}$ with no reflection, so that we could easily extrapolate $L_{2keV}$ assuming a simple power-law spectrum. However, in \citetalias{Liu_zhu+2016:XMM-XXL} the reflection component was also included in the calculation of $L_{2-10keV}$, as it is usually observed both in low-$z$ \citep[e.g.][]{Nandra+2007:Sey_iron} and high-$z$ \citep[e.g.][]{Baronchelli+2018:refl} spectra \citep[see the average-AGN model in][]{Bucnher+2014:BXA_agnrefmodel}. Therefore, we consistently excluded from the analysis all the sources with a significant reflection component: given the high errors of the typical $\log R$ fit in \citetalias{Liu_zhu+2016:XMM-XXL}\footnote{$R$ is the ratio of the normalization of the reflection component with respect to the power-law component.}, we included only sources in which the 16th percentile was $<-0.2$ and the 84th was $<0.5$. We note that \citetalias{Liu_zhu+2016:XMM-XXL} included in the fit also a scattering contribution from ionized material inside the angle of the torus \citep[see][]{Bucnher+2014:BXA_agnrefmodel}, although the fit normalizations are on the order of $10^{-4}$ with respect to the main power-law component.

\begin{figure}[tb]
	\centering
	\includegraphics[width=0.9\columnwidth]{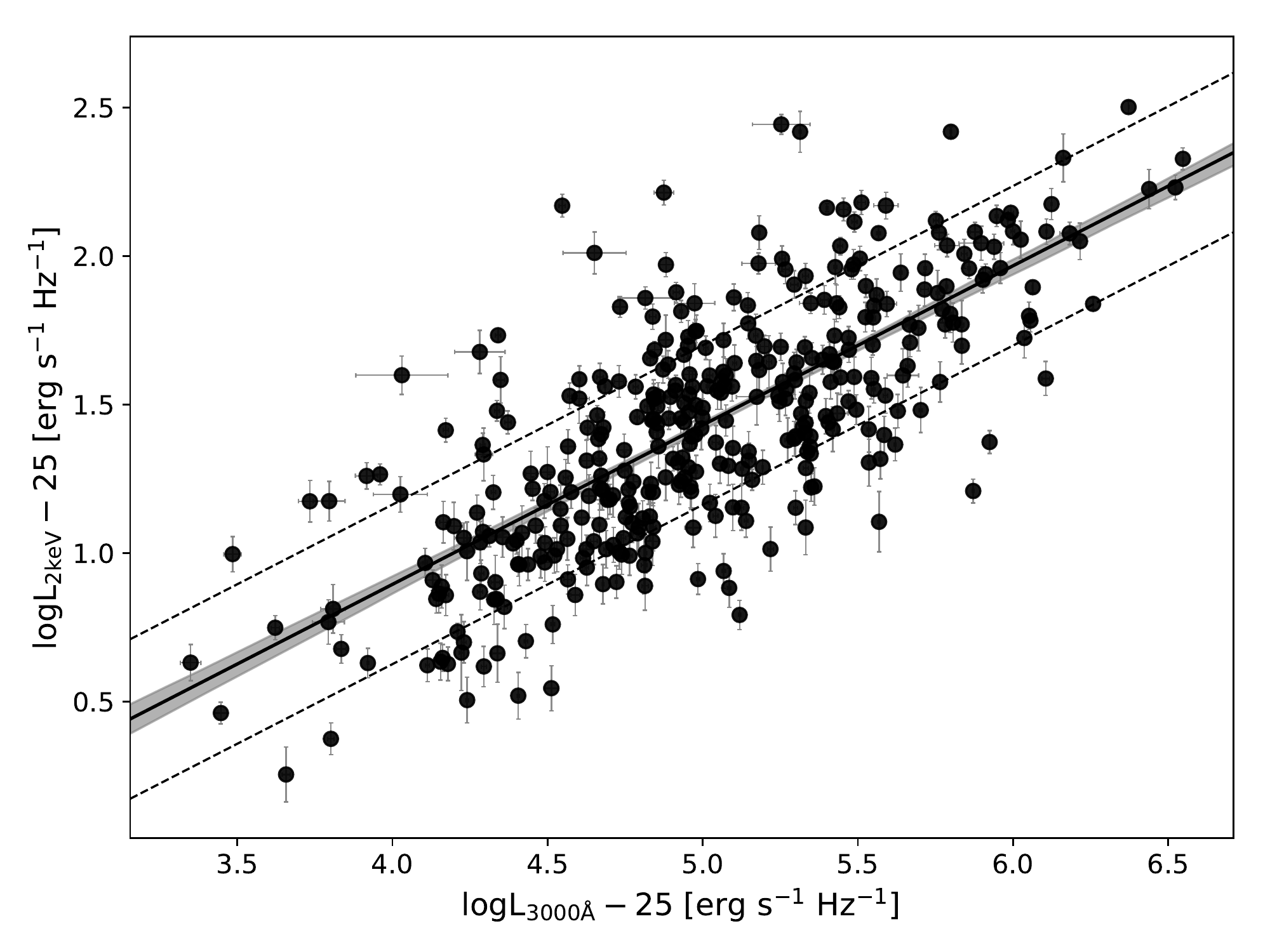}
	\caption{$L_{X}-L_{UV}$ relation of the 379 bright BLAGN of XMM-XXL. Monochromatic luminosity values are here scaled by 25~dex, to ease the comparison with recent works. The solid black line is the median regression line obtained with emcee, with the corresponding 16th and 84th percentiles represented with the shaded gray area. The dashed black lines show the intrinsic scatter around the median relation.}
	\label{fig:Lx_Luv_XMMXXL}
\end{figure}

The final cleaned subsample, to which we will refer as XMM-XXL, consists of 379 sources with observed $m$ (with median $\log m=8.7_{-0.5}^{+0.4}$), $L_{3000\AA}$, $\Gamma$ (with median $\Gamma=2.1\pm0.1$) and $L_{2keV}$. In Fig.~\ref{fig:Lx_Luv_XMMXXL} we show the $\log L_{X}=\widehat{\alpha}+\widehat{\beta}\log L_{UV}$ relation, with the best-fit linear regression given by:\begin{equation}\label{eq:best_fit2D}
\log L_{2keV} -25= (-1.25\pm 0.12)\,+(0.54\pm 0.02)\,(\log L_{3000\AA}-25)
\end{equation}
with intrinsic scatter $\sigma_{intr}=0.27\pm 0.01$. Linear regressions in two (or more) dimensions were performed with emcee \citep{Foreman-Mackey+2013:emcee}, accounting for uncertainties on all variables and an intrinsic scatter using the likelihood provided in \citet{Dagostini2005:fits}. The uncertainty in the independent variable(s) is propagated with the derivative $\partial Y/\partial X$ calculated in $X$ ($X_i$), equal to the slope coefficient(s) in the linear case \citep{DAgostini2003:}. The slope we measure is slightly flatter than what is quoted in the recent literature \citep[e.g.,][]{Lusso&Risaliti2016:LxLuvtight}, although we did not consider all the possible biases of flux-limited samples (however, see Appendix~\ref{sec:Lx_Luv_XXL}). For the main scope of this paper, it is sufficient to have a reference sample cleaned in accord with the physics described within the model.


\subsection{Methodology of the observational test}
\label{sec:method}

In our model $\dot{m}=\lambda_{edd}=L_{bol}/L_{edd}$, although we do not take as reference also $\dot{M}$ or $\lambda_{edd}$ from \citetalias{Liu_zhu+2016:XMM-XXL}: the former is interpolated from the mass and a monochromatic optical luminosity \citep{Davis&Laor2011:efficiency_Mdot}, while the latter depends on a disk-luminosity estimate via $L_{bol}$. Both approaches are based on standard-disk assumptions or calculations and using those values within our non-standard disk models would be an inconsistency. One can also estimate $L_{bol}$ applying bolometric corrections (BC) to the observed monochromatic optical-UV luminosities \citep{Richards+2006:qso_sed,Runnoe+2012:BC}, although the many uncertainties in play \citep{Krawczyk+2013:BC_caveats,KilerciEcer2018:BC_caveats} and the high scatter in the BCs \citep{Richards+2006:qso_sed,Lusso+2012:BCbothXandBband} discouraged us in relying on this approach. Then, for every source we iteratively obtain the $\dot{m}$ value yielding a model $L_{3000\AA}$ consistent with the observed one within its errors (typically $\sim0.01$\,dex). This approach is similar to the interpolation method put forward by \citet{Davis&Laor2011:efficiency_Mdot}, although we do it consistently for each different model, which is given by a choice of $\mu$, $\alpha_0$ and $f_{max}$. 

The methodology then consists in fixing $\mu$, $\alpha_0$ and $f_{max}$ (see Table~\ref{tab:model_param} for a summary on the model's parameters), which will be referred to as the model choice, within a discrete 3D grid in $\mu=[0,0.2,0.4,0.5,0.6,0.8,1]$, $\alpha_0=[0.02,0.2]$ and $f_{max}=[0.1,0.2,0.3,0.5,0.7,0.9,0.99]$. Then, we take as input $m$, $\Gamma$ and $L_{3000\AA}$ from the observed data, allowing us to solve the equations of the model for each source and compute  $\dot{m}$ and $L_{2keV}$ values (see Section~\ref{sec:model}). For each observed source of the reference sample, every model in the 3D grid can provide a mock entry for the $L_{X}-L_{UV}$. A proper comparison requires uncertainties to be assigned on the mocks, as the observed $m$, $\Gamma$ and $L_{3000\AA}$ come with their own measurement and systematic errors, where obviously the $\sim0.4-0.5\,$dex systematics in the mass estimates \citep[e.g.][and references therein]{Shen2013:QSOmass} play the dominant role. As it is mentioned above, mock $L_{3000\AA}$ values converge to the related observed quantities within their errors, hence we conservatively fixed the mock $\delta L_{3000\AA}$ at the 90th percentile of the uncertainty distribution in the observed $L_{3000\AA}$ (i.e. $\sim0.03\,$dex). In order to compute uncertainties for $\dot{m}$ and $L_{2keV}$, we ran each model 200 times on the same source, extracting the input values ($m$, $\Gamma$ and $L_{3000\AA}$) from a normal distribution with mean and standard deviation taken from the observed quantities and their errors. Then, the uncertainty on $\dot{m}$ and $L_{2keV}$ is taken from the dispersion of the 200 runs. 


\section{Results of the observational test}
\label{sec:results}

\begin{figure}[tb]
	\includegraphics[width=0.9\columnwidth]{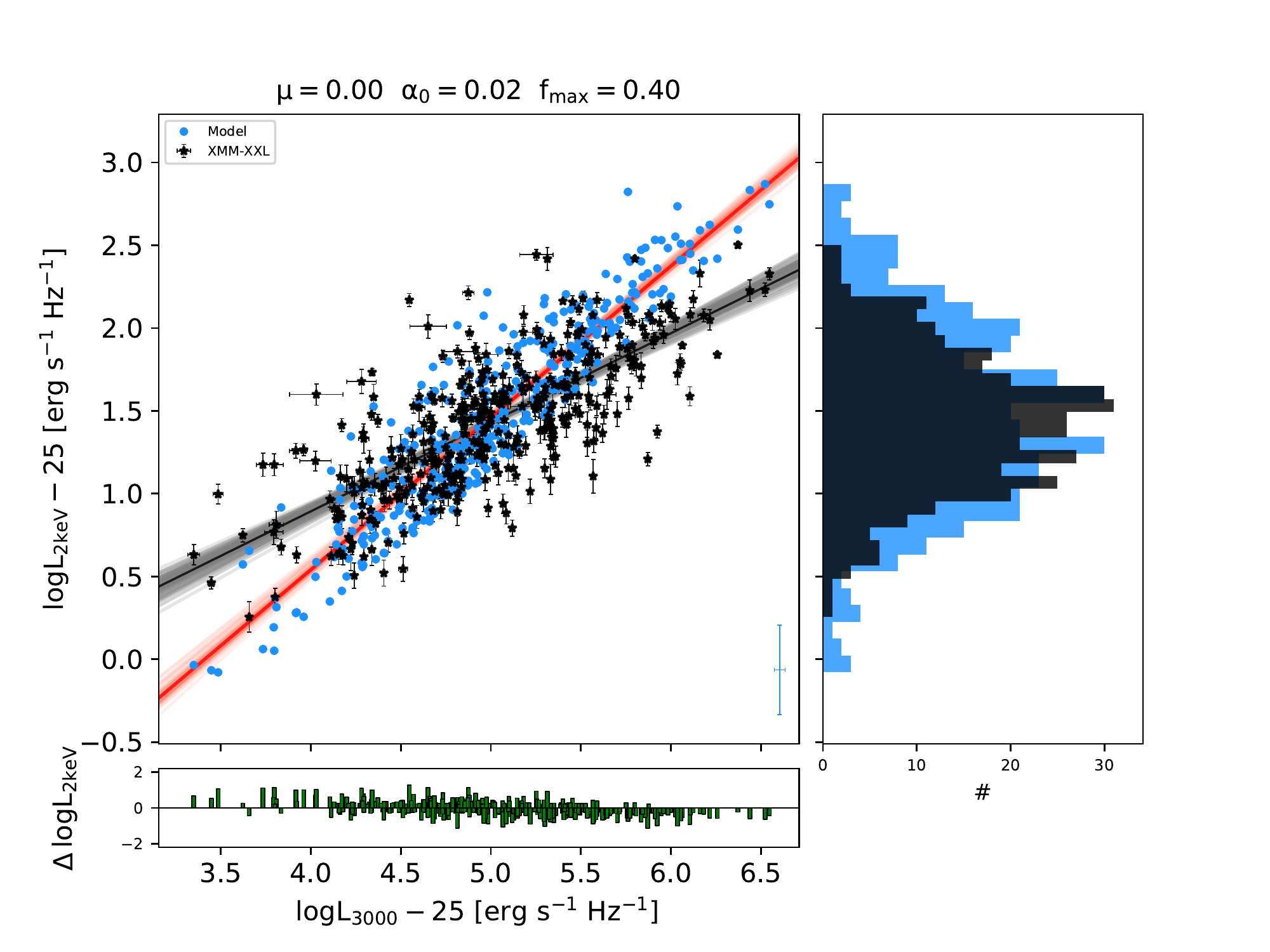}	
	\\
	\includegraphics[width=0.9\columnwidth]{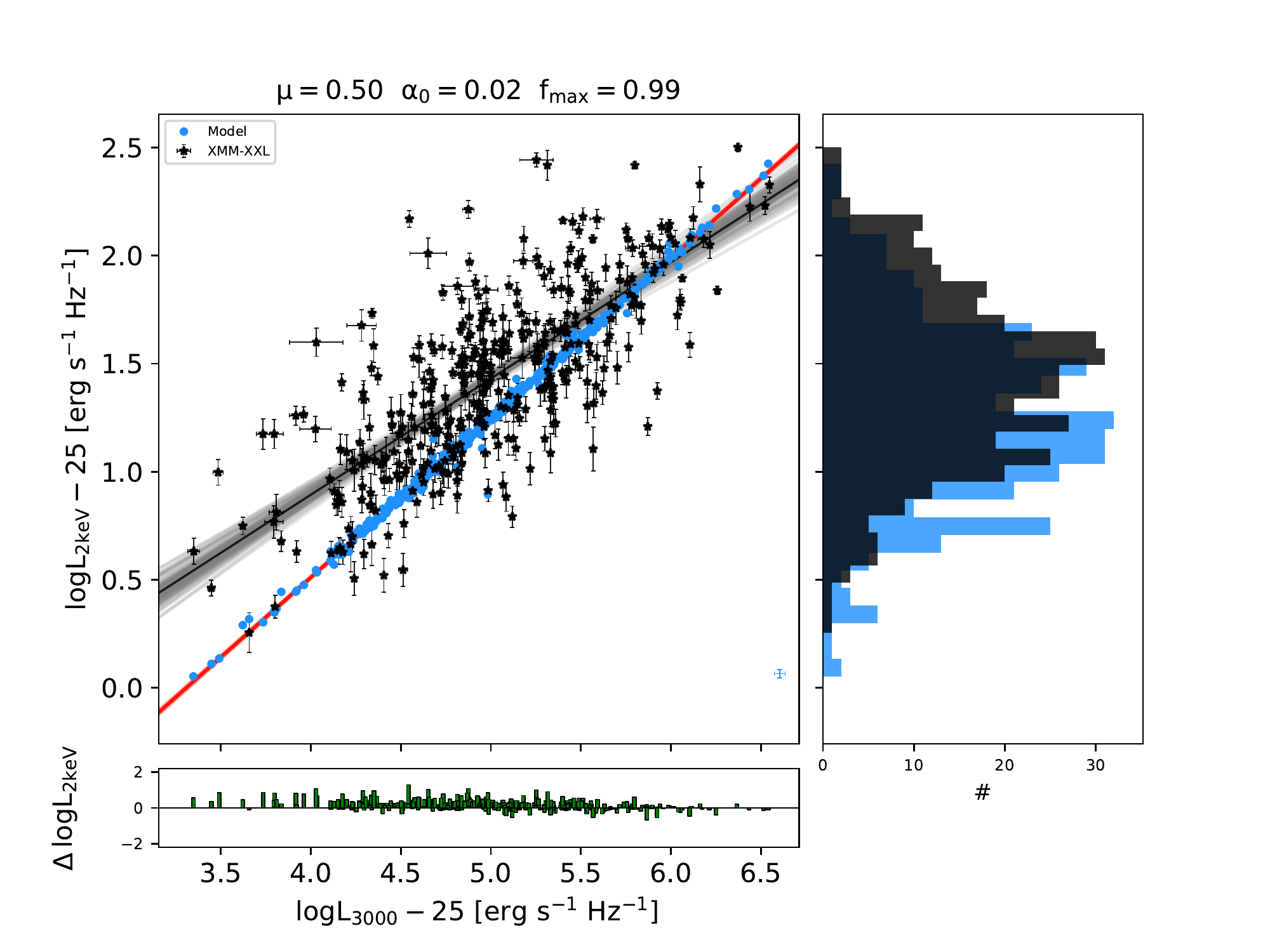}
	\\	
	\includegraphics[width=0.9\columnwidth]{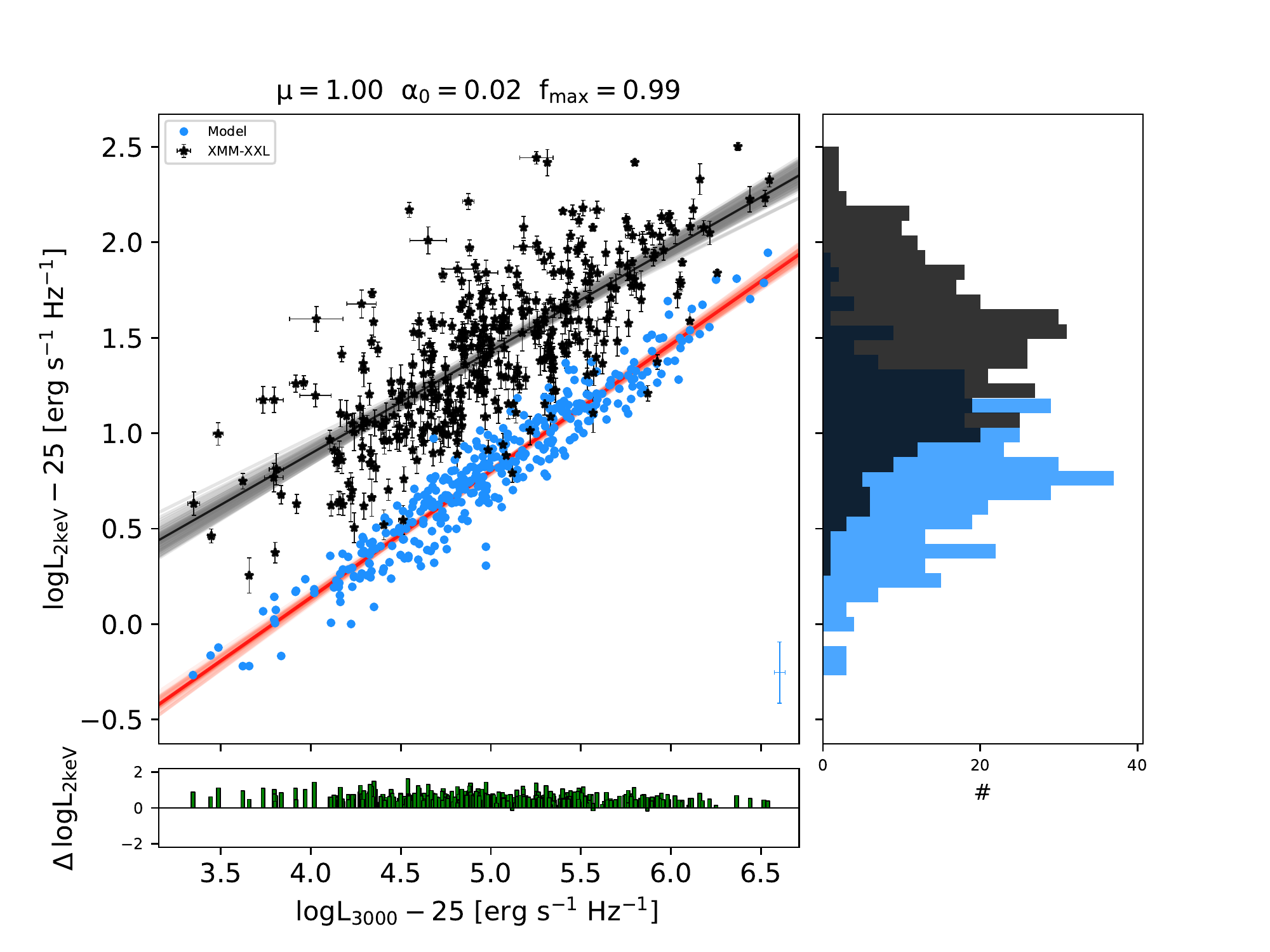}	
	\caption{The central panel of each image shows an example of the $L_{X}-L_{UV}$ relation for both XMM-XXL (black stars) and the model (blue dots), in which the choice of $\mu$, $\alpha_0$ and $f_{max}$ is shown in the titles. The black and red solid lines are randomly drawn from the posterior distributions of normalization and slope for XMM-XXL and the model, respectively, with the median regression line thickened. The bottom panels show the residuals given by the difference of observed and mock $\log L_{2keV}$ and the right panels show the related distributions. The errors on the model are show in the bottom right corner of the central panels.}
	\label{fig:mock_LxLuv}
\end{figure}

For all the models on the discrete 3D grid in the $\mu$, $\alpha_0$ and $f_{max}$ parameter space (see Section~\ref{sec:method} and Table~\ref{tab:model_param}), we fit the $L_X$-$L_{UV}$ distribution with a log-linear relation $\log L_{X}=\widehat{\alpha}+\widehat{\beta}\log L_{UV}$. Three examples are shown in Fig.~\ref{fig:mock_LxLuv} for $\mu$ corresponding to the known analytic viscosity prescriptions (see Section~\ref{sec:model}). Ideally, a model should reproduce the observed $L_{X}-L_{UV}$ in both normalization and slope. However, we can start decomposing the problem in two parts: a good match in the normalization ($\widehat{\alpha}$) would state that globally, for a given optical-UV luminosity distribution, the modeled corona emission was strong enough (see Section~\ref{sec:results_norm}); instead, if the slope ($\widehat{\beta}$) is matched, then the model accurately describes how the coronal strength varies from lowly- to highly-accreting sources (see Section~\ref{sec:results_slope}). Moreover, as it can be seen from the examples in Fig.~\ref{fig:mock_LxLuv}, our models come with their one intrinsic scatter, given by different $m$ and $\Gamma$ at a fixed $\dot{m}$. This provides precious insights on the nature of the total observed scatter (see Section~\ref{sec:results_scatter}). 

\subsection{The normalization of the $L_{X}-L_{UV}$}
\label{sec:results_norm}

First, we investigate how well the mocks reproduce the data normalization along the vertical axis of the $L_{X}-L_{UV}$. To do so, we define a score for the goodness of match:\begin{equation}
\label{eq:r2score}
r^2_i=1-\frac{\sum\limits_{i}\Big(y_{data,i}-y_{mock_i}\Big)^2}{\sum\limits_{i}\Big(y_{data,i}-<y_{data}>\Big)^2}
\end{equation}
The $r^2$\,score is computed drawing 1000 random samples from the observed $\log L_{2keV}$ within their errors (i.e. $y_{data,i}$), and 1000 random regression lines from emcee's chains on the mock (i.e. $y_{mock_i}$). Then, the median and the 84th-16th inter-quantile range are quoted from the resulting distribution of 1000 $r^2_i$\,scores. Negative scores indicate the data are poorly reproduced by the model; an $r^2=0$ would be obtained by a constant value corresponding to the mean of the observed $\log L_{2keV}$ distribution. We can put a quality threshold and keep all the models that yield a positive score. 

\begin{figure}[tb]
	\centering
	\includegraphics[width=0.9\columnwidth]{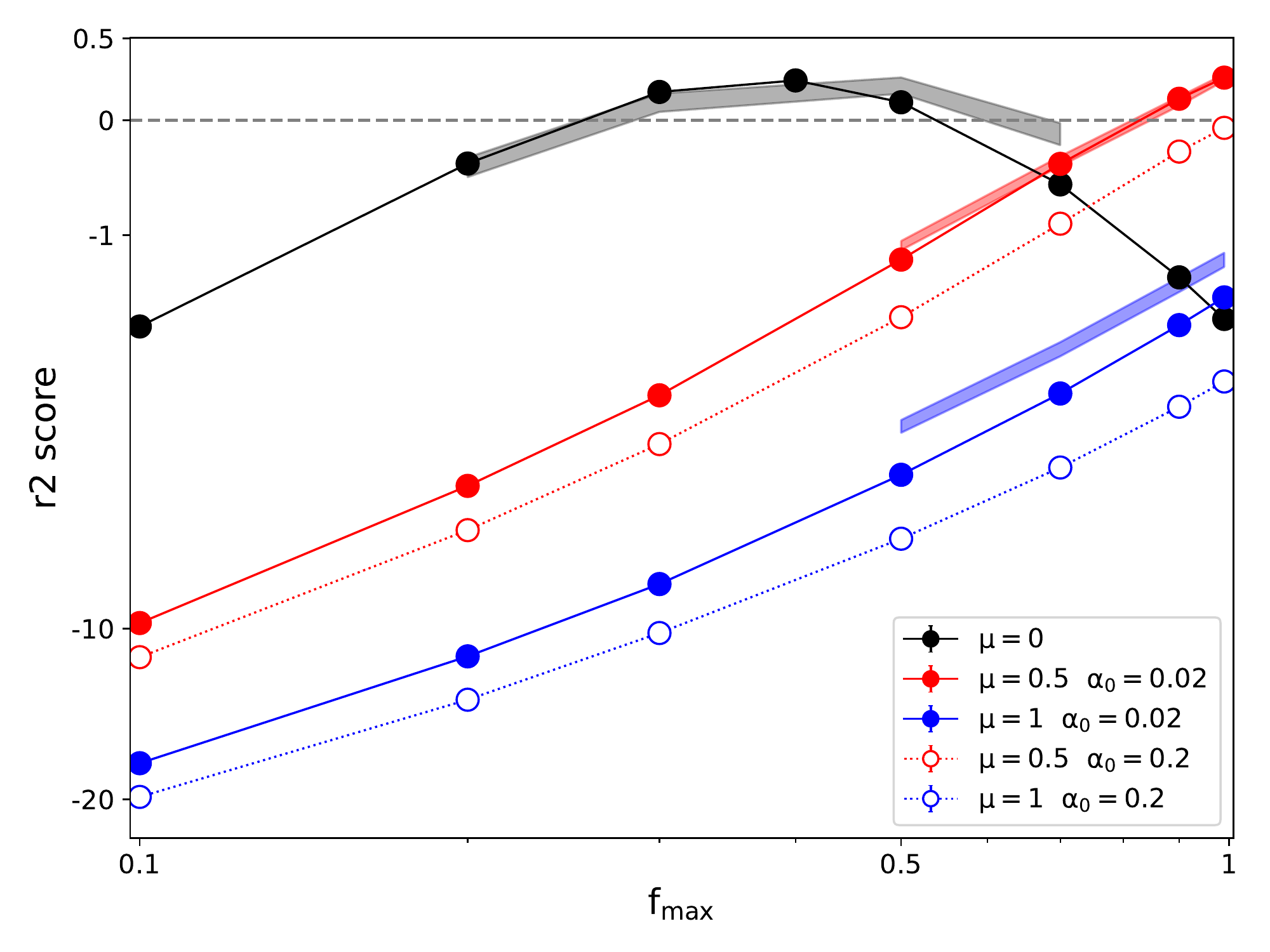}
	\caption{$r^2$\,score, representing the goodness of match between XMM-XXL and mocks (see text), as a function of $f_{max}$. Models with $\mu=0$, 0.5, 1 are color coded in black, red and blue, respectively. The additional dependency on $\alpha_0$ is represented with varying line-types as shown in the legend, when present (it is absent for $\mu=0$). A good match is represented with a score greater than zero. The points include the uncertainties in the score values. The shaded areas represent the results obtained applying the same methodology on a different sample (RM-QSO, Liu et al. in prep), fixing $\alpha_0=0.02$ and using the same colors. 
	}
	\label{fig:r2score_plot}
\end{figure}

The $r^2$\,score as a function of $f_{max}$ is shown in Fig.~\ref{fig:r2score_plot}, where the choice of $\mu$ is color coded and the additional dependency on $\alpha_0$ is represented with varying line-types (it is minor or absent, as in $\mu=0$). We show for simplicity only values of $\mu$ corresponding to the known analytic viscosity prescriptions (see Section~\ref{sec:model}). The other values used would accordingly show intermediate results. For each viscosity law there is a preferred $f_{max}$, that fixes the maximum coronal strength in a model. 

\begin{figure}[tb]
	\centering
	\includegraphics[width=0.99\columnwidth]{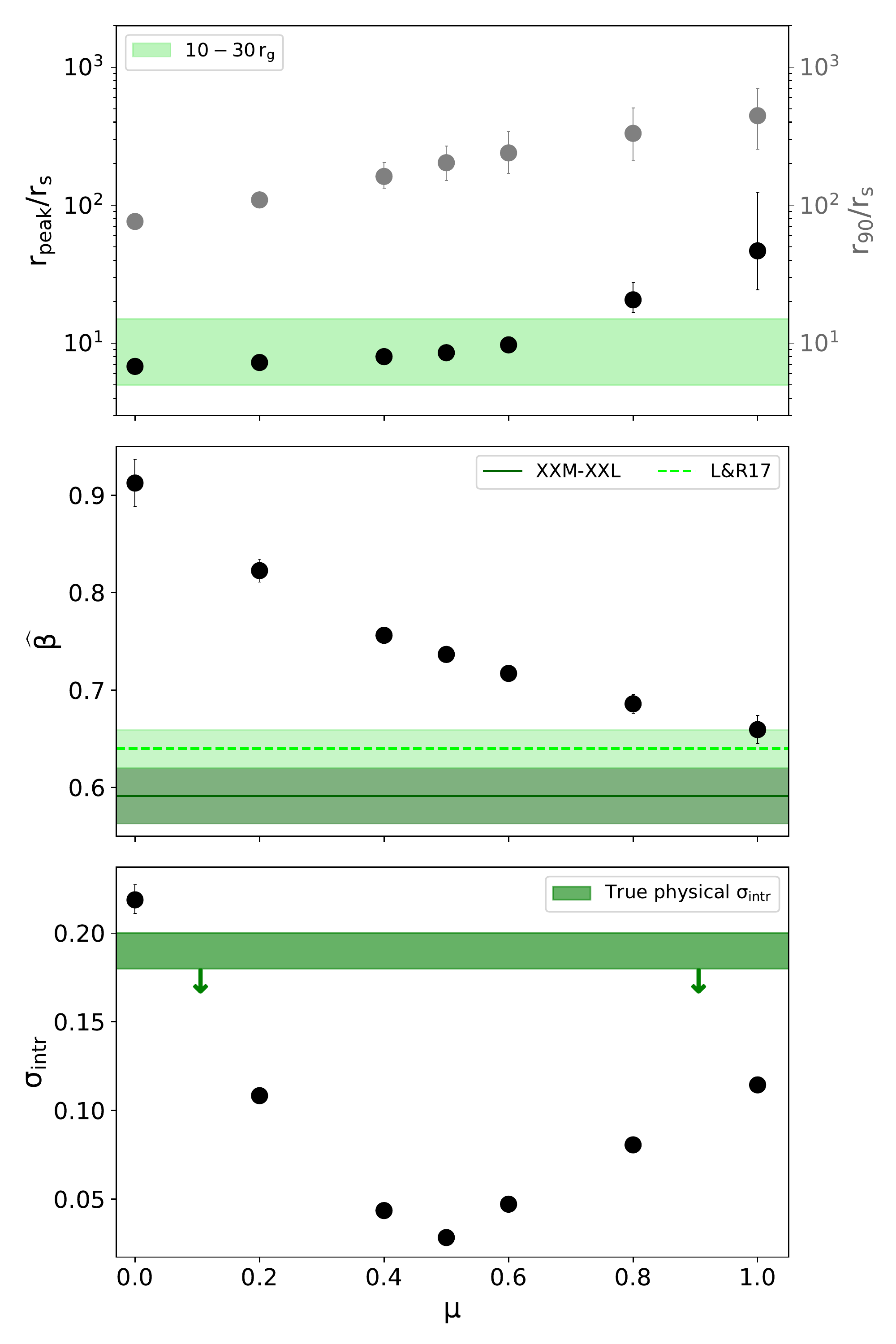}
	\caption{Top panel: the $2\,$keV-emission $r_{peak}$ ($r_{90}$) as a function of $\mu$ is represented in black (gray). The green shaded area qualitatively shows the inner radii, where the bulk of X-ray emission is supposed to come from according to X-ray reverberation and micro-lensing. For increasing $\mu$, the X-ray emission profile peaks at larger radii. Middle panel: for increasing $\mu$ the models obtain a slope of the $L_{X}-L_{UV}$ closer to the observed one. The dark-green area represents the reference slope of the cleanest XXM-XXL (Appendix~\ref{sec:Lx_Luv_XXL}), while the light-green refers to the slope quoted in \citetalias{Lusso&Risaliti2017:toymodel}. Bottom panel: intrinsic scatter of the mock $L_X-L_{UV}$ relations as a function of $\mu$. The green area represents a tentative upper limit of the true scatter \citep{Lusso&Risaliti2016:LxLuvtight,Chiaraluce+2018:dispandvariab_Lx_Luv}, that is only due to the physical properties of AGN. For simplicity, all panels show only the results obtained with a single $f_{max}$, corresponding to the highest $r^2$-score (e.g., Fig.~\ref{fig:r2score_plot}), and fixed $\alpha_0=0.02$.}
	\label{fig:rpeak_and_slope_vs_mu}
\end{figure}

Models with higher $\mu$ need higher normalization $f_{max}$, since they have a comparably weaker X-ray emission, in accord with their lower $\text{<}f\text{>}_{median}$ (see Fig.~\ref{fig:f_Lx_profiles} and Section~\ref{sec:radial_profiles}). Nonetheless, the law correspondent to $\mu=1$ does not produce adequately strong coronae even with $f_{max}=0.99$ and can be ruled out (see right image in Fig.~\ref{fig:mock_LxLuv}). Furthermore, $\mu=1$ produces a radially flatter X-ray emission profile (see bottom panel of Fig.~\ref{fig:f_Lx_profiles}), in contrast with observations that hint for coronae peaking in the inner radii \citep[e.g.][]{Mosquera+2013:lensedQSO_corona,Reis+2013:size_reverb_micro,Wilkins+2016:modeling_Xreverb}. We explore this behavior more quantitatively in the top panel of Fig.~\ref{fig:rpeak_and_slope_vs_mu} showing how the radius of the annulus at which the $2\,$keV emission peaks ($r_{peak}$, or at which it is $90\%$ of the total, $r_{90}$) varies with $\mu$: as $\mu$ increases, most of the corona emission comes from annuli placed at larger and larger radii. 

We also verified that our results do not depend on the sample adopted as reference. We performed the same analysis with the RM-QSO sources (Liu et al. in prep; \citealp{Shen+2018:RM-QSO}), on which a similar analysis was performed and on which we applied compatible cleaning criteria and methodology, as described in Sections~\ref{sec:sample} and~\ref{sec:method}. The results are shown in the $r^2$\,score plot (Fig.~\ref{fig:r2score_plot}) with shaded areas, color coded for $\mu$ in the same way and using only $\alpha_0=0.02$. There is generally a good agreement between the two samples, suggesting that our results are not dependent from the different data used.

It is worth stressing that the $f_{max}$ value at which each $\mu$ (possibly) matches the observed normalization is degenerate with the assumptions on the accretion efficiency and on the product $\eta\,(1-a_{disk})$. Namely, higher accretion efficiencies and/or a higher fraction of the coronal emission beamed away from the disk would increase the normalization of the $L_X$-$L_{UV}$ relation, and shift all curves of Fig.~\ref{fig:r2score_plot} to the left. This will be further examined in Sections~\ref{sec:impact_efficiency} and~\ref{sec:impact_DC}.

\subsection{The slope of the $L_{X}-L_{UV}$}
\label{sec:results_slope}

Figure~\ref{fig:r2score_plot} allows us to track the models (i.e. combinations $\mu$, $\alpha_0$ and $f_{max}$, see Table~\ref{tab:model_param}) that broadly reproduce the normalization $\widehat{\alpha}$ of the observed $L_{X}-L_{UV}$. Nonetheless, obtaining the correct normalization is simply a weighting exercise of the energetic outputs of the disk and the corona. It is the slope that carries the exact information on how the disk-corona interplay changes across the different accretion regimes of bright radiatively-efficient AGN (see Section~\ref{sec:prediction_lxluv}). This would require a precise knowledge of the true slope of the the $L_{X}-L_{UV}$. The observations suggest a value around $\approx 0.6$ (\citealp{Lusso&Risaliti2016:LxLuvtight}; \citetalias{Lusso&Risaliti2017:toymodel}; our Appendix~\ref{sec:Lx_Luv_XXL}) and we can acknowledge this value as reference. Our methodology, however, can be regarded as data-independent, and it would applicable even if future works will update the current knowledge on the exact value of the slope.

In the middle panel of Fig.~\ref{fig:rpeak_and_slope_vs_mu} we show how the modeled slope of the of the $L_{X}-L_{UV}$ gets closer to the observed one for increasing $\mu$ (i.e. for more damped radial $f$-profiles), for a fixed $\alpha_0=0.02$ and using only the $f_{max}$ corresponding to the highest $r^2$-score. This is because models with increasing $\mu$ have higher logarithmic scatter in $f(r)$, meaning that going from lowly- to highly-accreting sources the span in $\text{<}f\text{>}_i$ is larger, with high-$\dot{m}$ objects having comparably weaker X-ray emission with respect to low-$\dot{m}$ companions (see Section~\ref{sec:prediction_lxluv}). We show this for $\mu=0$, 0.5 and 1, respectively\footnote{The distributions of mock $\dot{m}$ are very similar across the models, with median values (and related 16th and 84th percentiles) of $0.16_{0.04}^{0.69}$, $0.15_{0.05}^{0.65}$ and $0.14_{0.04}^{0.59}$ for $\mu=0$, 0.5 and 1, respectively. The tails include Eddington or even super-Eddington sources. We note that the uncertainty on the modeled $\dot{m}$, propagated through the ones in the observations, is as large as $\approx0.65\,$dex.}:
\begin{equation}\label{eq:fmean_m_mdot}
\begin{split}
\log\text{<}f\text{>}&= \log f_{max} \\
\log\text{<}f\text{>}&= (-1.12\pm 0.24)\,-(0.15\pm 0.02)\,\log\dot{m}\,\\
&+\,(0.05\pm 0.03)\,\log m \\
\log\text{<}f\text{>}&= (-1.82\pm 0.36)\,-(0.27\pm 0.03)\,\log\dot{m}\,\\
&+\,(0.07\pm 0.04)\,\log m \\
\end{split}
\end{equation}
where the steepest dependency from $\dot{m}$ is obtained for larger $\mu$. 

This test points in the same direction as the evidence of an X-ray bolometric correction increasing with the accretion rate \citep[e.g.][]{Wang2004:hotdisccorona_constraints,Vasudevan+Fabian2007:kbol_1,Vasudevan+Fabian2009:kbol_2,Lusso+2010:alphaOX,Young+2010:alpha_ox_kbol}, although we refrain to compare this observable with our regressions \citep[e.g.][]{Wang2004:hotdisccorona_constraints,Cao2009:coronamodel,Liu2009:disc-corona_investigated,You2012:model_disc_corona,Liu+2012:corona_model_highL,Liu+2016:structure_spec_corona_highL}, due to the many more uncertainties in play when deriving bolometric luminosities in comparison to the quantities entering in the $L_{X}-L_{UV}$ (see the discussion in Section~\ref{sec:method}). 


\subsection{The scatter of the $L_{X}-L_{UV}$}
\label{sec:results_scatter}

The observed scatter of the $L_{X}-L_{UV}$ for the sample used in this work is $\sigma_{intr}=0.27\pm 0.01$ (Section~\ref{sec:sample}). As a matter of fact, this value represents an upper limit to the intrinsic dispersion inherent to the physics of the system, as the observed scatter is affected by a combination of instrumental and calibration issues, UV and X-ray variability, non-simultaneity of the multi-wavelength observations. A lot of effort has been put into trying to quantify as accurately as possible all these contaminants \citep[e.g.][and references therein]{Vagnetti+2013:variab_X_uv,Lusso2018:instrum_on_variab}, with claims that the intrinsic scatter in the $L_X$-$L_{UV}$ relation is smaller than  $\lesssim 0.18-0.20$ \citep{Lusso&Risaliti2016:LxLuvtight,Chiaraluce+2018:dispandvariab_Lx_Luv}. Any successful model should be able to reproduce such a low scatter.

From the examples of mock $L_{X}-L_{UV}$ relations plotted in Fig.~\ref{fig:mock_LxLuv}, it can already be seen that our models come with their one intrinsic scatter. In our methodology (Section~\ref{sec:method}), the modeled $\dot{m}$ was tuned to the observed $L_{3000\AA}$, hence the intrinsic scatter of the mock $L_{X}-L_{UV}$ relations is simply the dispersion of the modeled $L_{2keV}$, at a given $\dot{m}$, due to different $m$ and $\Gamma$. We show this more quantitatively in the bottom panel of Fig.~\ref{fig:rpeak_and_slope_vs_mu}. The models dispersion varies with $\mu$ because changing the viscosity law induces a different logarithmic scatter in $f(r)$ (see Fig.~\ref{fig:f_Lx_profiles}) and it also affects the distance (in gravitational radii) from which the bulk of the $L_{2keV}$ is coming (see top panel of Fig.~\ref{fig:rpeak_and_slope_vs_mu}). The resulting $\sigma_{intr}$ of the models is likely a complex combination of these (and possible more) factors. All the models, with the exception of $\mu=0$, lie below the available observational constraints \citep{Lusso&Risaliti2016:LxLuvtight,Chiaraluce+2018:dispandvariab_Lx_Luv} of $\lesssim 0.18-0.20$. This is another successful prediction of our model (see Section~\ref{sec:prediction_lxluv}).

\subsection{A complete picture: the slope-normalization plane of the $L_{X}-L_{UV}$}
\label{sec:results_slope_norm}

\begin{figure}[tb]
	\centering
	\includegraphics[width=0.99\columnwidth]{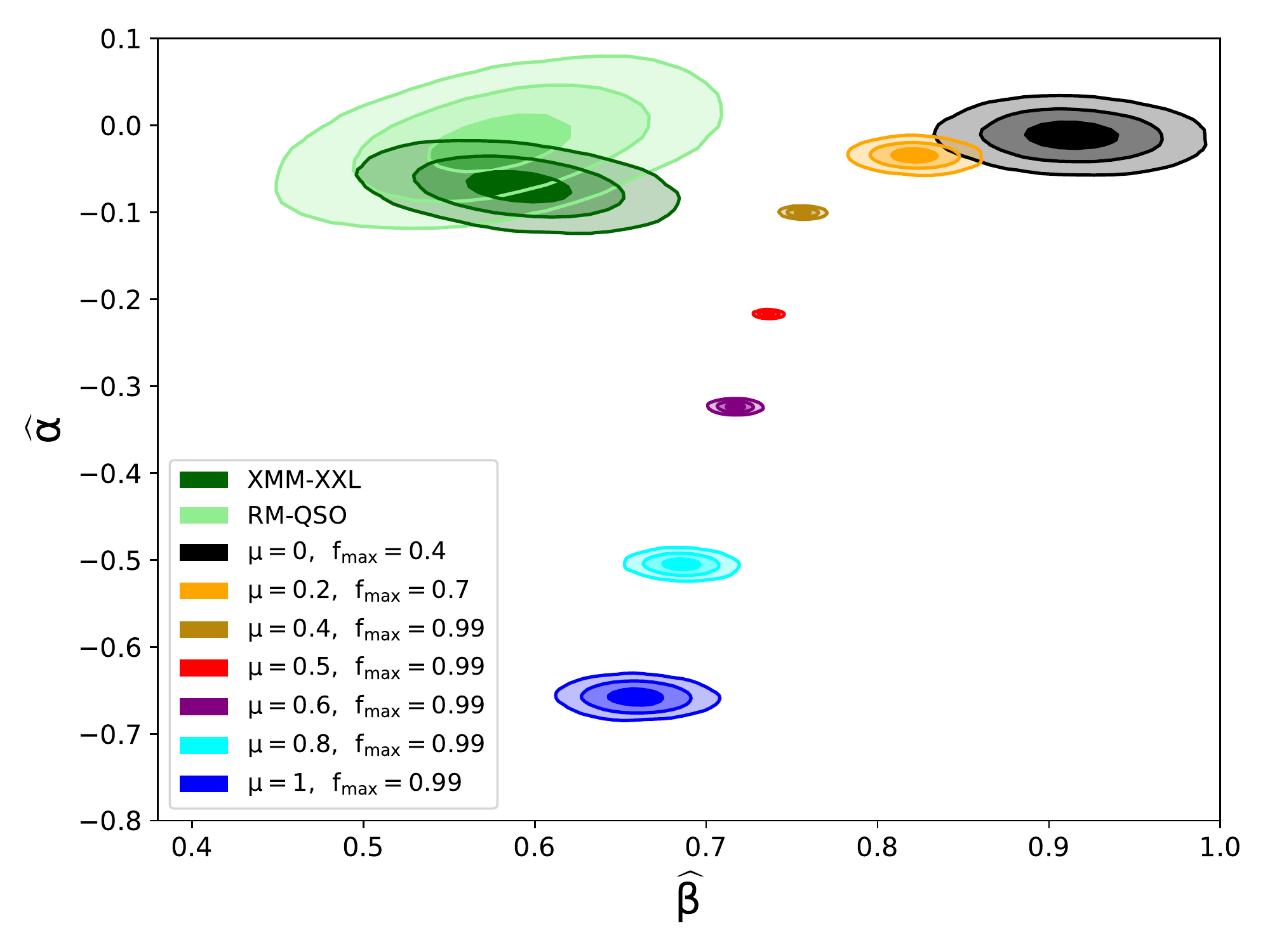}
	\caption{1-, 2- and 3-sigma contours of the emcee regressions in the slope-normalization ($\widehat{\beta}-\widehat{\alpha}$) plane of the $L_{X}-L_{UV}$ for both data and models, normalizing all $L_{X}$ and $L_{UV}$ to the corresponding median values of XMM-XXL. Dark green contours are related to the cleanest XMM-XXL sample (Appendix~\ref{sec:Lx_Luv_XXL}) and the light green ones to the RM-QSO sources. The contour of the models are color coded for $\mu=[0,0.2,0.4,0.5,0.6,0.8,1]$, as shown in the legend. For simplicity, we report for each $\mu$ only results obtained with a single $f_{max}$, corresponding to the highest $r^2$-score, and fixed $\alpha_0=0.02$. Models that reproduce the observed slope $\widehat{\alpha}$ are also the ones that show weaker coronae (lower normalization $\widehat{\alpha}$).}
	\label{fig:norm_slope_plane}
\end{figure}

In the previous Sections, we decomposed the match in either normalization or slope to have a better understanding on how our disk-corona models can relate to the observed $L_{X}-L_{UV}$. However, the goal would be to have a model that can fully encompass these observables. Hence, in Fig.~\ref{fig:norm_slope_plane} we display 1-, 2- and 3-sigma contours in the slope-normalization plane ($\widehat{\beta}-\widehat{\alpha}$) of the $L_{X}-L_{UV}$ for both data and models. All regressions were performed with emcee normalizing both $L_{X}$ and $L_{UV}$ to the median value of XMM-XXL. The data contours are related to the cleanest XMM-XXL version (Appendix~\ref{sec:Lx_Luv_XXL}) and to the RM-QSO sources\footnote{\label{fnote:balmerRM}XMM-XXL luminosities were obtained in \citetalias{Liu_zhu+2016:XMM-XXL} including a Balmer continuum component in the fit \citep[refer to][]{Shen&Liu2012:fit_QSO_shengroup}, although for the RM-QSO this component was switched off \citep{Shen+2018:RM-QSO}. For consistency, a rigid shift of $-0.12\,$dex was applied to the RM-QSO $L_{3000\AA}$ \citep{Shen&Liu2012:fit_QSO_shengroup} for obtaining the contours displayed in Fig.~\ref{fig:norm_slope_plane}.}. Model contours are shown for $\mu=[0,0.2,0.4,0.5,0.6,0.8,1]$ using a single $f_{max}$, corresponding to the highest $r^2$-score (e.g., Fig.~\ref{fig:r2score_plot}) for each $\mu$, and a fixed $\alpha_0=0.02$, for simplicity.

Fig.~\ref{fig:norm_slope_plane} shows that models reproducing the observed slope, namely the ones with higher $\mu$ (as in middle panel of Fig.~\ref{fig:rpeak_and_slope_vs_mu}), are also the ones that show weaker coronae (lower normalization $\widehat{\alpha}$) and overly extended $L_{2keV}$-emission (i.e. higher $r_{peak}$ and $r_{90}$, top panel of Fig.~\ref{fig:rpeak_and_slope_vs_mu}). 

\subsection{The 3D plane: $L_{X}$ vs $L_{UV}$ vs $m$}
\label{sec:results_3D}

As shown by \citetalias{Lusso&Risaliti2017:toymodel}, the $L_{X}-L_{UV}$ relation for AGN is rather a three-dimensional problem, with the mass (or its proxy given by the full-width half-maximum of broad emission lines) playing a significant role as well. The observed $L_{X}-L_{UV}-m$ plane from XMM-XXL can be fit by:
\begin{equation}\begin{split}\label{eq:Lx_Luv_m_XXM}
\log L_{2keV} -25 &= (-0.91\pm 0.13)\,+(0.39\pm 0.03)\,(\log L_{3000\AA} -25) \\
&+(0.23\pm 0.04)\,(\log m -7)
\end{split}\end{equation} and the mock $L_{X}-L_{UV}-m$ from models with $\mu=0$, 0.5 and 1, respectively:\begin{equation}\begin{split}\label{eq:Lx_Luv_m_models}
\log L_{2keV} -25 &= (-3.49\pm 0.15)\,+(1.08\pm 0.03)\,(\log L_{3000\AA} -25) \\
&-(0.27\pm 0.03)\,(\log m -7) \\ \\
\log L_{2keV} -25 &= (-2.41\pm 0.15)\,+(0.73\pm 0.01)\,(\log L_{3000\AA} -25) \\
&+(0.013\pm 0.004)\,(\log m -7) \\ \\
\log L_{2keV} -25 &= (-2.28\pm 0.08)\,+(0.57\pm 0.02)\,(\log L_{3000\AA} -25) \\
&+(0.14\pm 0.02)\,(\log m -7)
\end{split}\end{equation}The comparison in the 3D plane states that the exact dependency is not obtained by any of the models, with $\mu=1$ being the closest in qualitatively retrieving the coefficients for $L_{3000\AA}$ and $m$. We note that the mass is taken from the observations, thus this mismatch states that the luminosities in the model do not depend on the mass in the correct way.

\subsection{The impact of the accretion efficiency}
\label{sec:impact_efficiency}

\begin{figure}[tb]
	\centering
	\includegraphics[width=0.99\columnwidth]{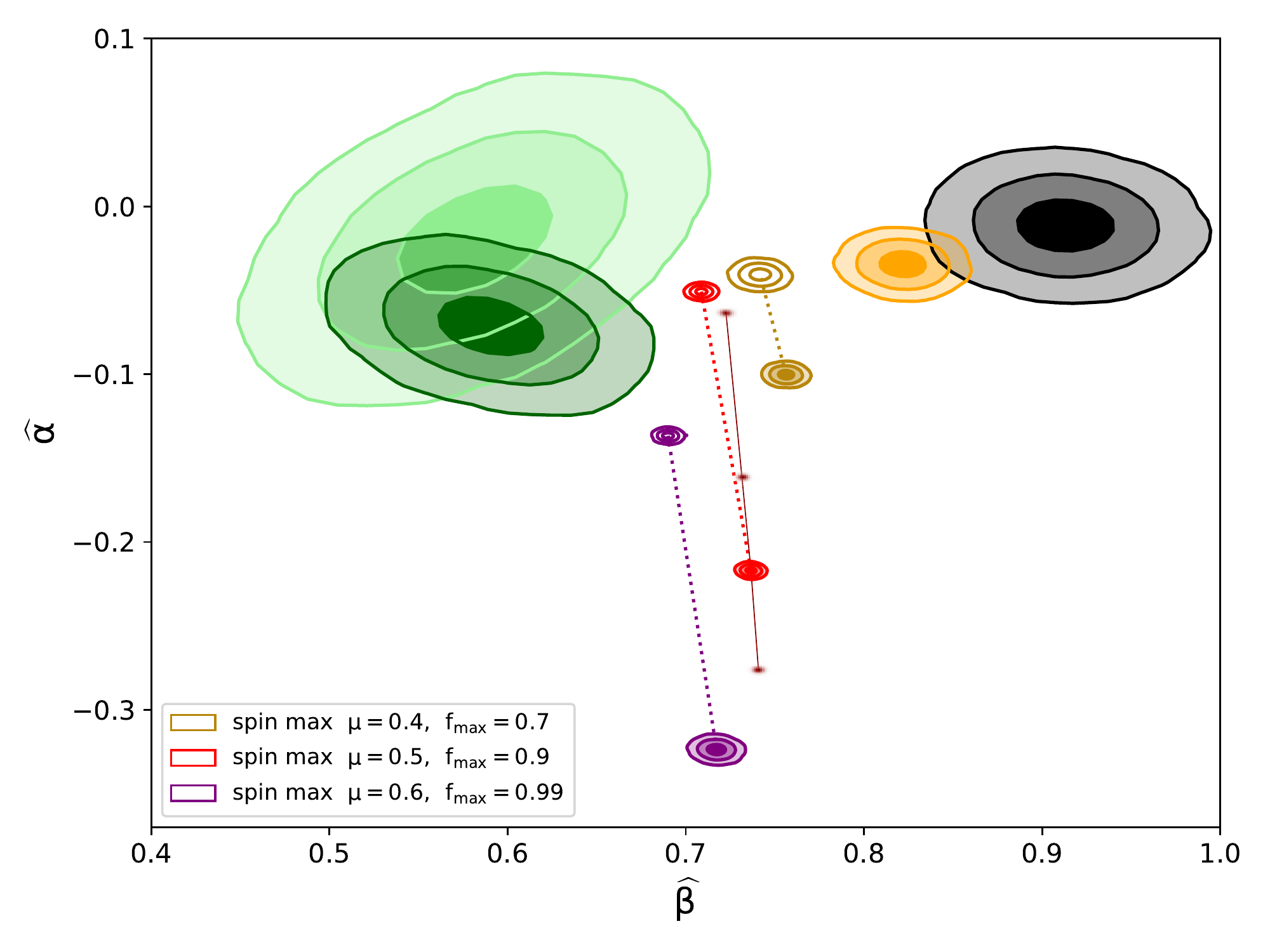}
	\caption{Same as Fig.~\ref{fig:norm_slope_plane}, with the addition of empty contours for $\mu=0.4$, 0.5 and 0.6 (color coded in the legend) obtained with maximally spinning black holes (i.e. with $\epsilon_0=0.3$ and $r_0=1.24r_g$). The dashed lines connect them to the non-spinning analogous realizations. Dark-red density spots represent the location of the center of different contours of the standard $\mu=0.5$ case, in which the only difference is the adoption of $\eta$ (downward scattering component) varying among 0.4, 0.5 and 0.6, going from higher to lower $\widehat{\alpha}$, respectively.}
	\label{fig:discussion_efficiency}
\end{figure}

Throughout this work we adopted an efficiency $\epsilon_0=0.057$, typical of non-rotating black holes \citep[e.g.][]{Shapiro2005:BH_growth}, for simplicity. Nonetheless, a high spin seems to be preferred to model the blurred relativistic iron line, detected both in the local Universe \citep{Nandra+2007:Sey_iron,REynolds2013:localAGN_FEspin} and up to $z\sim4$ \citep[e.g.][]{Baronchelli+2018:refl}. Moreover, flux-limited samples are known to be biased in preferentially detecting high-spinning black holes \citep{Brenneman+2011:high-spin_bias,Vasudevan+2016:high-spin_bias2}, simply because they are brighter than their non-rotating analogous \citep[see][]{Reynolds2019:obs_spin}.

Then, we tested the model using maximally-spinning black holes, with radiative efficiency 0.3 and ISCO down to $r_0=1.24r_g$ \citep{Thorne1974:BHs}. This has a major impact on the normalization axis of the $L_X-L_{UV}$.
Everything else in the source being equal, in a spinning black hole matter can be accreted down to smaller distances with respect to their non-rotating companions, thus the accretion power in the system is much higher. As a matter of fact, changing the radiative efficiency has an impact on the numerical equation that regulates f(r): for the same $m$ and $\dot{m}$ and $r>3$ the values of $f$ is higher, and the transition radius between $P_{rad}$- and $P_{gas}$-dominated regions moves at lower radii. This self-consistently affects the disk equations via the $(1-\tilde{f})$ factor (see Appendix~\ref{sec:app_model}), hence the surface temperature is decreased at higher radii, where most of the disk emission at $3000\AA$ comes from. Then, the modeled $\dot{m}$ value needed to match the observed $L_{3000\AA}$ is higher (see Section~\ref{sec:method}) and, consequently, $L_{2keV}\propto fQ_+$ is higher. 

In Fig.~\ref{fig:discussion_efficiency} we show the model contours in the correlation slope-normalization plane computed for both low and high radiative efficiency, for $\mu=0.4$, 0.5 and 0.6 only.

\begin{figure}[tb]
	\centering
	\includegraphics[width=0.75\columnwidth]{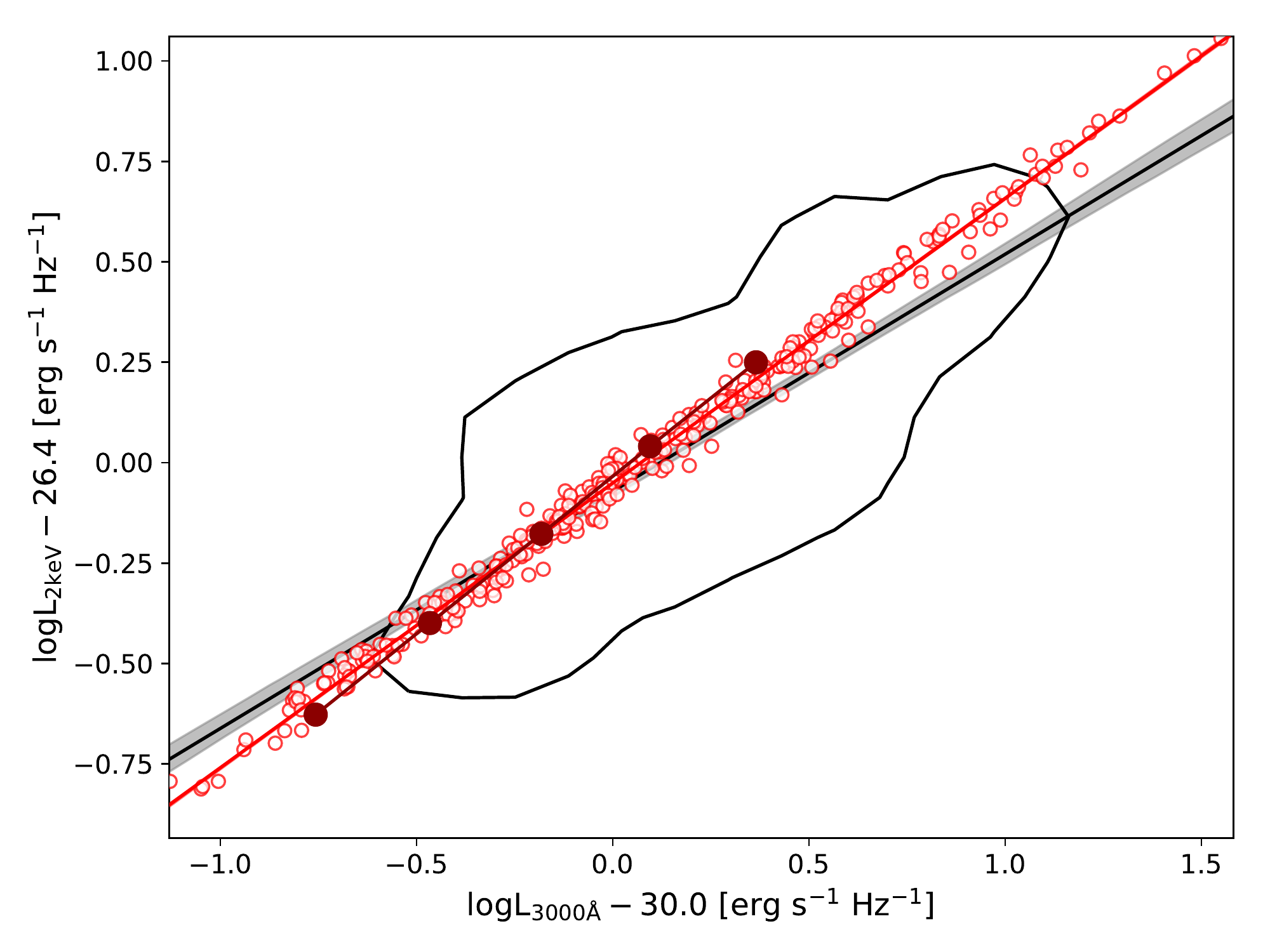}
	\caption{$L_{X}-L_{UV}$ relation for the high-spin model with $\mu=0.5$, $\alpha_0=0.02$ and $f_{max}=0.9$ (empty red points, corresponding to the empty red contour in Fig.~\ref{fig:discussion_efficiency}), with the red line showing best fit slope from emcee. The connected filled points (dark red) show the single-mass trend ($\log m =8.7$) for varying accretion rate ($0.03, 0.07, 0.17, 0.42, 1 $). For a comparison, the black contour shows where the data lie in the plane, with the related best-fit slope (black line).}
	\label{fig:discussion_efficiency_lxluv}
\end{figure}

Interestingly, maximally-spinning sources yield a better match with the data contours, in particular for the viscosity law $\mu=0.5$, with $f_{max}=0.9$. For instance, Fig.~\ref{fig:discussion_efficiency_lxluv} shows how the data and this high-spin model compare in the $L_{X}-L_{UV}$ plane. We want to stress that using only a maximum spin for all sources is an extreme measure, but since the (unknown) observed spin distribution is likely dominated by high-spin values \citep{Reynolds2019:obs_spin}, model contours of a more realistic diverse population of high-spinning sources would be closer to the high-efficiency ones in Fig.~\ref{fig:discussion_efficiency} rather than to the spin-zero case. We also note that, even if the modeled coronae would be somewhat weaker using a realistic spin distribution, with respect to the maximum-spin case, the model with $\mu=0.5$ can still be realized with a higher $f_{max}=0.99$. Thus, we speculate that the new empty red contours in Fig.~\ref{fig:discussion_efficiency} consist in a fair approximation of a realistic high-spin population model. The tension with the observed $L_{X}-L_{UV}$ would be significantly relaxed.  

\subsection{The impact of the downward scattering component}
\label{sec:impact_DC}

The results shown in Fig.~\ref{fig:r2score_plot} are also degenerate with the assumptions on the value of the product $\eta\,(1-a_{disk})$, that is on the assumed downward component of the X-ray emission ($\eta$) and on the disk albedo. The adopted value of $\eta=0.55$ is typical for anisotropic Comptonization in a plane-parallel corona \citep{Haardt+Maraschi1993:twophase2}, although it is unclear how much it would change in different geometries or prescriptions. In a patchy corona \citep{Haardt+1994ApJ:patchy} $\eta$ would unlikely part significantly from the one in the slab case. The only major difference would rather involve the transmission or absorption by the corona of the radiation reflected by the disk. However, we conservatively excluded from the reference sample adopted in the observational test all the sources with a non-negligible reflection component detected (see Section~\ref{sec:sample}), allowing us to avoid its complicated modeling. In an outflowing corona \citep[e.g.][]{Beloborodov+1999:HE_accr,Malzac+2001:dyn_coronae}, the ratio between the downward and the upward flux  decreases with the bulk velocity of the corona \citep[e.g.][]{Janiuk+2000:outflow_corona}. We tried to quantify possible offsets in the $\widehat{\beta}$-$\widehat{\alpha}$ plane due to different values of $\eta$, ranging from $0.6$ (slightly enhanced downward scattering) to $0.4$ (reduced downward scattering, roughly approximating an outflowing corona with $\beta_{bulk}\approx0.1-0.2$, e.g. \citealp{Janiuk+2000:outflow_corona}). We show this in Fig.~\ref{fig:discussion_efficiency} for the $\mu=0.5$ case, with dark-red density spots ($\eta=0.4$, 0.5 and 0.6 from higher to lower $\widehat{\alpha}$, respectively). Changing the downward component by $\Delta\eta\sim0.1$ induces a significant offset of $\approx0.1\,$dex in $\widehat{\alpha}$ and a minor change in $\widehat{\beta}$.

\section{Discussion}
\label{sec:discussion}

The $L_{X}-L_{UV}$ relation has been studied for decades \citep[starting with the better-known $\alpha_{OX}$ parameter,][]{Tananbaum+1979:alphaOX}, its robustness used for bolometric estimates \citep[e.g.][]{Marconi+2004:locSMBH,Hopkins+2007:bolQLF,Lusso+2010:alphaOX} and recently even for cosmology \citep{Risaliti&Lusso2015:Hubble_diagram,Risaliti&Lusso2018:cosmo2}. Nonetheless, there is currently no solid and exhaustive physical explanation for it. In Section~\ref{sec:prediction_lxluv} we outlined the qualitative predictions of our model and in Section~\ref{sec:results} we obtained that concordance with current data can be obtained with a modified viscosity prescription in the accretion flow ($\mu=0.5$), provided the spin of the sources is high. Here, we briefly discuss whether other competing analytic disk-corona models succeed or not and then we try to investigate the impact of the assumptions in our model on the results.

\subsection{Comparison with other models}

\begin{figure}[tb]
	\centering
	\includegraphics[width=0.75\columnwidth]{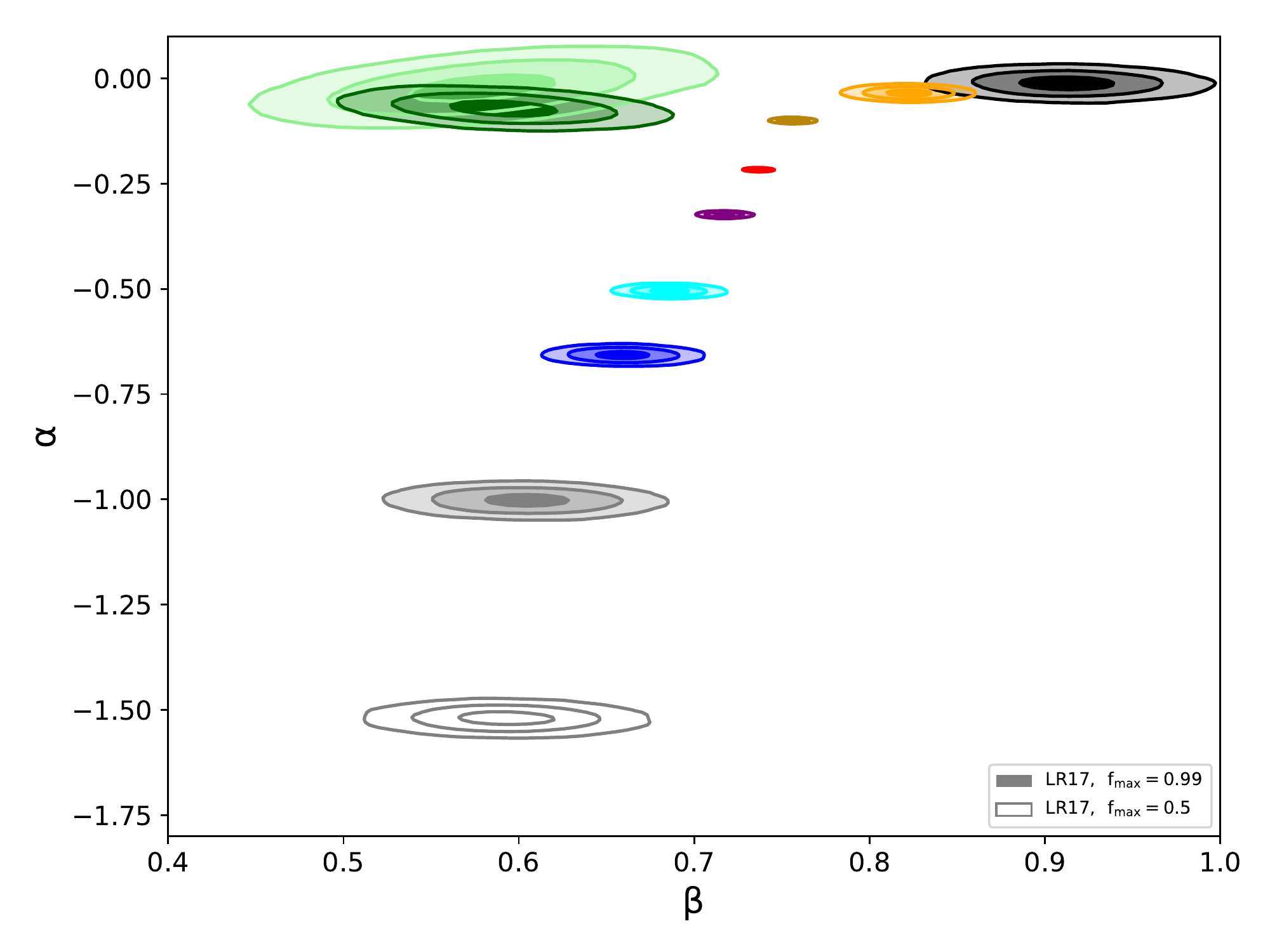}
	\caption{Same as Fig.~\ref{fig:norm_slope_plane} with the addition of a reproduction of \citetalias{Lusso&Risaliti2017:toymodel}'s toy-model as gray contours.}
	\label{fig:discussion_LR}
\end{figure}

\citetalias{Lusso&Risaliti2017:toymodel} tried to explain this relation with a very simplified, but effective, toy-model. Most of their assumptions are in common with our work (see Section~\ref{sec:limits_model}), although our treatment is more complete and physically motivated. The assumption of the MRI amplifying the magnetic field to a lesser extent in $P_{rad}$-dominated regions (\citealp{Blaes&Socrates2001:Prad_MRI,Turner+2002:Prad_MRI_simul}; \citetalias{Merloni2003:model}) is taken to the extreme with a step function for the $f$-profile: all the accretion power is emitted by the disk in $P_{rad}$-dominated regions (i.e. $f(r_{rad})=0$), whereas it is equally distributed between disk and corona in $P_{gas}$-dominated regions (i.e. $f(r_{gas})=0.5$). The resulting predicted slope and normalization of the $L_{X}-L_{UV}$ are claimed to be consistent with the observations. The former can be confirmed by our analysis, as their $f(r)$ step-function is nothing but an extremely damped $f(r)$ beyond $\mu=1$, whose mock slope of the $L_{X}-L_{UV}$ was the closest to the observed one. In the latter case, their match in normalization might be an involuntary artifact: with respect to the power transferred to the corona $f$, the observed luminosity is roughly halved if a downward scattering component is included (i.e. $f (1-\eta)$, with $\eta\approx 0.5$). 
We verified this running our model with $\mu=0$ and $\alpha_0=0.02$, forcing $f=0$ in the $P_{rad}$-dominated region and fixing both $f=0.50$ and $f=0.99$ in $P_{gas}$-dominated radii. In Fig.~\ref{fig:discussion_LR} we show the related contours in the $\widehat{\beta}-\widehat{\alpha}$ plane along with our results of Fig.~\ref{fig:norm_slope_plane}. This confirms that their step $f$-profile results in a slope consistent with the observed value, albeit producing overly weak coronae (too low normalization in the $\widehat{\beta}-\widehat{\alpha}$). Hence, their toy-model does not reproduce the $L_{X}-L_{UV}$. Moreover, the X-ray emission from their toy-model inevitably peaks at the transition radius between $P_{rad}$- and $P_{gas}$-dominated regions. Indeed, their model with $f_{gas}=0.99$ yields $r_{peak}=142_{\,51}^{\,438}$ and $r_{90}=790_{\,445}^{\,1490}$ (i.e. produces extremely extended coronae).

\citet{Kubota+Done2018:model_lx_luv} coupled an outer standard disk with an inner warm Componising region, that produces the soft X-ray excess, and an innermost hot corona for the hard X-ray continuum. Their model fits remarkably well the broadband continua of three sources spanning a wide range of accretion rates. They also claim to reproduce the observed $L_{X}-L_{UV}$, using both the regression line and data points from \citetalias{Lusso&Risaliti2017:toymodel}, although only displaying all the possible sources modeled within a grid of $m=10^6-10^{10}$ and $\dot{m}=0.03-1$ (their Fig. 7 and 8). Nonetheless, first-order normalization matches, even with $m$ and $\dot{m}$ spanning within typical values, can be misleading. A more conclusive test would be, as we do, to match mock and data sources one by one.


\subsection{Further assumptions and theoretical uncertainties}
\label{sec:limits_model}

Only models with $\mu\lesssim0.4$ are able to reproduce the observed normalization within the range of possible $f_{max}$ values, whereas for $\mu\gtrsim0.5$ they are off by $\gtrsim0.1-0.2\,$dex along the normalization. In Sections~\ref{sec:impact_efficiency} and~\ref{sec:impact_DC} we showed how a higher accretion efficiency and/or a different downward scattering component may affect our results in the slope-normalization plane. Their impact would be significant and can possibly ease the tension between data and models: high-spin black holes and/or moderately outflowing coronae would be consistent with the observations. We now try to investigate some other simplifications of our model, likely to have a minor or less quantifiable effect on our conclusions. 

\subsubsection{Soft X-ray excess and thermal instability}

The XMM-XXL $L_{2keV}$ value was interpolated from the $L_{2-10\,keV}$ fit in \citetalias{Liu_zhu+2016:XMM-XXL} after excluding sources with high reflection fraction (Section~\ref{sec:sample}). The impact of the soft X-ray excess component can be considered negligible in that energy range, thus data points in the $L_{X}-L_{UV}$ are likely not contaminated. However, our models do not include a soft X-ray excess generation mechanism, the monochromatic $L_{2keV}$ being extracted from a power-law spectrum within $0.1-100\,$keV. If a significant fraction of the power dissipated in the corona is actually used by a different mechanism producing the observed soft-excess, namely from a warm corona \citep[e.g.][]{Petrucci+2018:warm_corona,Kubota+Done2018:model_lx_luv,Middei+2018:2coronamodel}, the mock $L_{2keV}$ would be overestimated to an unclear extent. Nonetheless, if the soft X-ray excess is produced by blurred relativistic reflection \citep[e.g.][]{Crummy+2006:blurred_refl,Garcia+2018:blurred?}, the influence of this component on our analysis would have been excluded with our selection criteria (Section~\ref{sec:sample}). 

In Section~\ref{sec:model}, we briefly addressed the disk-instability problem (see Fig.~\ref{fig:f_Lx_profiles}, top panel) and despite the local stabilizing effect of the iron bump in the opacities, disks with $\mu=0$ (e.g., \citetalias{Shakura+Sunyaev1973:accretion}) and 0.5 (e.g., \citetalias{Merloni2003:model}) are globally unstable in $P_{rad}$-dominated regions. An intriguing question may be whether the unstable regions in the disk are responsible for generating the soft-excess, possibly within inhomogeneous flows \citep[e.g.][]{Merloni+2006:inhom_discs}. As a matter of fact, the higher $\dot{m}$ the wider the region where $P_{rad}$ dominates and the higher the soft-excess strength \citep[e.g.][]{Boissay+2016:soft_excess_hardX}. Nonetheless, a more thorough investigation of this scenario is beyond the reach of this paper.

\subsubsection{Magnetically-dominated disks}

In our model the stress tensor is dominated by Maxwell stresses as confirmed by simulations \citep[e.g.,][]{Hawley+1995:Bstresses,Sano+2004:MHD_accrdisc,Minoshima+2015:MRI_gaspress}, although the magnetic pressure is bound to be only a fraction of the product $P_{gas}^{\mu}P_{tot}^{1-\mu}$ via $\alpha_0$ at the mid-plane. However, there are theories postulating disks that are $P_{mag}$-dominated also in the denser regions \citep[e.g.][and references therein]{Begelman+Silk2017:magn_elevated} and not only in the upper layers \citep[e.g.][]{Miller&Stone2000:corona_form}, possibly solving a few long-standing issues of the standard accretions disk theory \citep{Dexter+Begelman2019:Bdisk_variab}. Simulations indeed showed that $P_{mag}$ can become an important competitor in supporting the disk vertically \citep{Bai+Stone2013:accr_disc_MRI_corona,Salvesen+2016:netpoloidalfield}, although heavily depending on the strength of the net vertical magnetic field, the origin of which is not fully understood, yet. If this imposed net vertical field is small (if $\beta_0=P_{tot}/P_{mag}>>1$), the buoyant escape of the toroidal component, amplified by MRI, is faster than its creation and a disk-corona system consistent with our model is formed. However, the evidence of disks that are magnetically-dominated even at the mid-plane is supported by \citet{Jiang+2019:subEdd_disks}, that recently performed a global 3D radiation-MHD simulation of two sub-Eddingtion flows. The structure of their simulated disks is significantly different from the standard \citetalias{Shakura+Sunyaev1973:accretion} model and reaches a complexity that our simplified prescriptions are not able to grasp. On the other hand, these simulations could not produce spectra and luminosities, yet. We here rely on the assumption that the energetics of $P_{mag}$-dominated disks are not significantly different from standard thin disks at radii larger than $\sim10 r_g$ \citep[e.g., see][]{Sadowski2016:Bthisdisks}. 



\subsubsection{Winds and outflows}

In order to see if any known broad absorption line (BAL) quasars were present in our sample, we cross-matched the XMM-XXL catalog (\citetalias{Liu_zhu+2016:XMM-XXL}; \citealp{Menzel+2016:XXM-XXL_boss}) with SDSS-DR12 \citep{Paris+2017:SDSS-DR12}, that flagged 29580 BAL QSO after visual inspection. Only two sources among the 379 used in our analysis were flagged, although they were both assigned zero indexes in the common metrics used for a more quantitative measurement of the BAL properties \citep{Paris+2017:SDSS-DR12}. Hence, our sample has no contaminations from known BALs, although we can investigate the possible impact of un-modeled wind-dominated objects on our work. For instance, \citet{Nomura+2018:linedrivW} recently developed a disk model compensating for the mass-loss rates of UV-driven winds, while consistently adjusting the temperature and emission of the underlying disk. They referred to a future work for a more complete modeling of the inner radii and the hard X-ray emission, but the influence on $L_{3000\AA}$ values seems already significant, provided $\dot{m}\gtrsim0.5$. Since winds appear to act only from moderate to Eddington $\dot{m}$, neglecting their presence would have an impact on the modeled $L_{X}-L_{UV}$ slope. The wind carries away kinetic energy reducing the disk emission accordingly, thus for a given observed high $L_{3000\AA}$, our no-wind model would underestimate $\dot{m}$ for the possible outflowing sources contaminating our sample.

\subsubsection{The larger-than-predicted disk argument}

One of the most studied issues of the standard \citetalias{Shakura+Sunyaev1973:accretion} disk model is that observed sizes appear to be larger than expected at optical-UV wavelengths, using both microlensing effects \citep[e.g.][]{Morgan+2010:microlensQSO,Blackburne+2011:microl_fl_rat,Jimenez-Vicente+2012:QSO_microl} and flux variability lags across multiple bands in the so-called reprocessing scenario \citep[e.g.][]{Edelson+2015:swift_reverb_size,Fausnaugh+2016:delays_size,Fausnaugh+2018reverb_2sey,Jiang+2017:lags_panstarrs,Cackett+2018:reverb_BLR?,McHardy+2018:lags_caveats}, in which often a compact X-ray emitting region (e.g., a lamppost corona) irradiates the disk inducing light-travel lags in the UV-optical bands. However, even combining all these results discordant with the theoretical predictions is not trivial \citep{Kokubo2018:caveats_lags}, particularly if different techniques are used \citep[see][]{Moreno+2018:variab_techn,Vio+2018:CARMAissues}. What is more, there are also numerous studies finding consistency with the sizes predicted by the standard \citetalias{Shakura+Sunyaev1973:accretion} theory \citep[e.g.][]{McHardy+2016:lags_ss?,Mudd+2018ApJ:sizes-ss73?,Yu+2018:sizes_standard,Edelson+2018:swift_reverb_conststent,Homayouni+2018:sizes_consistent}, thus we do not consider necessary to use the larger-than-predicted argument to abandon all the standard prescriptions, yet.

\subsubsection{No-torque inner boundary}

For convenience, we adopted the no-torque condition with the stress vanishing at the inner edge. However, the presence of magnetic torques \citep{Gammie1999magnetic_torque,Agol+Krolik2000:magnetic_torque} would increase the disk effective temperature and the $Q_+$ emissivity in the innermost radii \citep{Agol+Krolik2000:magnetic_torque,Dezen+2018:nonzero_torque} and, if applied to the disk only, it would cause instead a drop in the fraction $f$ \citep{Merloni&Fabian2003:coronaGR}. Without a proper MHD treatment, it is unclear how the modeled $L_{2keV}\propto f Q_+$ would be affected, and consequently the $L_X-L_{UV}$ slope.

\subsubsection{The vertical structure}

Our model does not properly treat the vertical structure of the disk. The effective temperature is obtained from $T_{eff}\propto T_{mid}/\tau^{1/4}$, where $\tau=h\, \rho \,\kappa$ assumes constant $\rho$ and $\kappa$ along the scale-height. Even keeping the approximation of a constant $\rho$, $\kappa$ should change self-consistently with the decrease in temperature. A more thorough modeling of the disk vertical structure in supermassive black holes was presented by Hubeny and collaborators, taking into account both scattering processes and free-free and bound-free opacities \citep{Hubeny+2000:nonLTEdiscs,Hubeny+2001:part2_compton}. Their model also share some of our limits (e.g., stationary disk, $\alpha$-prescription, no-torque boundary, vertical support from thermal pressure only), validating the comparison. The overall SED has lower (higher) fluxes at low (high) frequencies with respect to standard calculations, with the most significant impact on the modeling of the soft-excess \citep{Done+2012:soft_excess}. The computation of $L_{3000\AA}$ should be affected in a minor way, with a small overestimation on the order of a color correction \citep[e.g.][]{Done+2012:soft_excess}, that is either roughly constant or weakly depending on $m$ and $\dot{m}$ \citep[e.g.][]{Davis+2018:spec_hardening}. Our conclusions should not be significantly affected, although this would need to be improved for a proper SED modeling and time-lags predictions.

\section{Conclusions}
The gap between simulations and observations in AGN needs to be bridged and simplified, but motivated, analytic prescriptions still represent a powerful tool to explain the observed multi-wavelength scaling relations. For instance, the clear correlation observed between monochromatic logarithmic $L_X$ and $L_{UV}$ luminosities has been used for decades \citep[in the shape of the more-known $\alpha_{OX}$ parameter,][]{Tananbaum+1979:alphaOX} in many applications \citep[even for cosmology, e.g.][]{Risaliti&Lusso2015:Hubble_diagram,Risaliti&Lusso2018:cosmo2}. Despite this, 
a conclusive theoretical explanation for the observed correlation is still lacking. Being smaller than one, the observed slope indicates that, going from low- to high-accretion rate AGN, the X-ray emission increases less than the optical-UV emission. Any viable disk-corona model must be able to explain this.


In this work, we tested a self-consistent disk-corona model (Section~\ref{sec:model}, see also \citetalias{Merloni2003:model}) against the $L_X-L_{UV}$ relation.
We were able to identify the possible mechanism regulating the disk-corona energetic interplay, in terms of viscosity prescriptions (e.g., $\mu=0.5$) that naturally lead to an X-ray emission increasing less than the disk emission going to higher accretion rates (see Section~\ref{sec:prediction_lxluv}). 



We also put forward a quantitative observational test (Section~\ref{sec:observational_test}), using a reference sample of AGN (Section~\ref{sec:sample}) observed both in the (rest-frame) UV and in X-rays: taking from each source the observationally determined $m$, $\dot{m}$ and $\Gamma$ we were able to model an analogous mock object (Section~\ref{sec:method}) producing a set of mock $L_X-L_{UV}$. 
This allowed us to reach a deep understanding of the physics driving the slope, normalization and scatter of the $L_X-L_{UV}$ (see Section~\ref{sec:results}).




We find that if the black-hole population is assumed to be non-spinning, results from this test are inconclusive: the viscosity prescriptions reproducing the slope of the observed $L_X-L_{UV}$ relation, also produce overly weak coronae. 
Interestingly enough, the tension between the strength of the observed and modeled X-ray emission (i.e. in the normalization of the $L_X-L_{UV}$) can be significantly relaxed adopting a more realistic high-spinning black-hole population and/or with moderately-outflowing coronae. We tested the former case adopting the efficiency (and the inner orbit) of maximally-spinning black holes, in which matter is able to accrete further into the potential well, resulting in a much higher accretion power and, consequently, in much stronger coronae (Section~\ref{sec:impact_efficiency}). Moreover, if the spin is high the X-ray emission profile peaks closer to the black hole, in even better agreement with X-ray reverberation and microlensing studies \citep[]{Mosquera+2013:lensedQSO_corona,Reis+2013:size_reverb_micro,Wilkins+2016:modeling_Xreverb}. In particular, the disk-corona model testing maximally-spinning black holes with $\mu=0.5$ (i.e. magnetic stress proportional to the geometric mean of $P_{gas}$ and $P_{tot}$, e.g. see \citetalias{Merloni2003:model}), $f_{max}=0.9$, $\alpha_{0}=0.02$ (see Table~\ref{tab:model_param}) provides the best match with the observations (Fig.~\ref{fig:discussion_efficiency_lxluv}), although the modeled slope is still somewhat larger than the observed one (Fig.~\ref{fig:discussion_efficiency}). Going beyond this type of exercises, only 3D global radiation-MHD simulations will be able to better disclose the disk-corona physics \citep[e.g.][]{Jiang+2017:globalSUPEREDD,Jiang+2019:subEdd_disks}, provided a clearer way of approaching the observations will be reached.

\begin{acknowledgements}
    We thank the referee for his/her helpful comments. We thank Teng Liu for kindly providing the optical spectral slopes obtained in \citep{Liu_teng+2018:Xobs_type1AGN}. We are also grateful to Torben Simm for the RM-QSO data and to Elisabeta Lusso and Guido Risaliti for making available to us the data of \citetalias{Lusso&Risaliti2017:toymodel}. RA thanks Damien Coffey, Jacob Ider Chitham and Linda Baronchelli for insightful discussions. We acknowledge the use of the matplotlib package \citep{Hunter2007:matplotlib}.
\end{acknowledgements}

%
%
\bibliographystyle{aa} 
\bibliography{bibliography} 

\begin{appendix}
\section{Disc-corona equations}
\label{sec:app_model}

We report the equations for $h$, mid-plane $\rho$ (g\,cm$^{-3}$), $P$ (dyn\,cm$^{-2}$), $T$ (K, at the mid-plane) and the closure equation for $f$. In the radiation pressure dominated regime:
	\small
	\begin{flalign*}
		\rho=\rho_{const}\,
		k_0^{-\frac{4}{\mu+4}}[\alpha_0m]^{-\frac{4}{\mu+4}}
		[\dot{m}J(r)]^{\frac{2(3\mu-4)}{\mu+4}}
		r^{\frac{3(2-3\mu)}{\mu+4}}
		(1-\tilde{f})^{\frac{6(\mu-2)}{\mu+4}} &&
	\end{flalign*}\begin{flalign*}
		T=T_{const}\,k_0^{-\frac{1}{\mu+4}}[\alpha_0m]^{-\frac{1}{\mu+4}}
		[\dot{m}J(r)]^{\frac{2\mu}{\mu+4}}
		r^{\frac{3(2\mu^2-3\mu-2)}{2(2-\mu)(\mu+4)}}
		(1-\tilde{f})^{\frac{2\mu-1}{\mu+4}} &&
	\end{flalign*}\begin{flalign*}
		h=9.14\,\dot{m}J(r)(1-\tilde{f}) &&
	\end{flalign*}\begin{flalign*}
		P=P_{const}\,k_0^{-\frac{4}{\mu+4}}[\alpha_0m]^{-\frac{4}{\mu+4}}
		[\dot{m}J(r)]^{\frac{8\mu}{\mu+4}}
		r^{\frac{6(2\mu^2-3\mu-2)}{(2-\mu)(\mu+4)}}
		(1-\tilde{f})^{\frac{4(2\mu-1)}{\mu+4}} &&
	\end{flalign*}\begin{flalign}\label{eq:system_rad_mu}
		\frac{(2\alpha_0)^{1/\mu}-k_1^{2/\mu}f^{2/\mu}}{k_1^{2/\mu}f^{2/\mu}}=
		\tilde{C}\,k_0^{\frac{1}{\mu+4}}[\alpha_0m]^{\frac{1}{\mu+4}}
		[\dot{m}J(r)]^{\frac{8}{\mu+4}}
		r^{-\frac{21}{2(\mu+4)}}
		(1-\tilde{f})^{\frac{9}{\mu+4}} &&
	\end{flalign}\normalsize where $k_0$ is the proportionality constant between the stress tensor and the magnetic pressure. The constant values depend on $\mu$ as follows:
\small
\begin{flalign*}\rho_{const}=\left(4.7\times10^{-68}\right)^{\frac{6(2-\mu)}{\mu+4}}
\left(5.5\times10^{48}\right)^{\frac{2(8-3\mu)}{\mu+4}}
\left(1.5\times10^{-23}\right)^{\frac{4\mu}{\mu+4}}\end{flalign*}
\begin{flalign*}T_{const}=\left(4.7\times10^{-68}\right)^{\frac{1-2\mu}{\mu+4}}
\left(5.5\times10^{48}\right)^{\frac{2(\mu^2-3\mu+2)}{(2-\mu)(\mu+4)}}
\left(1.5\times10^{-23}\right)^{\frac{\mu}{\mu+4}}\end{flalign*}\begin{flalign*}P_{const}=\frac{a}{3}\left(4.7\times10^{-68}\right)^{\frac{4(1-2\mu)}{\mu+4}}
\left(5.5\times10^{48}\right)^{\frac{8(\mu^2-3\mu+2)}{(2-\mu)(\mu+4)}}
\left(1.5\times10^{-23}\right)^{\frac{4\mu}{\mu+4}}\end{flalign*}\begin{flalign}
\tilde{C}=\left(4.7\times10^{-68}\right)^{\frac{-9}{\mu+4}}
\left(5.5\times10^{48}\right)^{\frac{-10}{\mu+4}}
\left(1.5\times10^{-23}\right)^{\frac{4}{\mu+4}}\end{flalign}\normalsize

The solutions for gas pressure dominated regions, that are independent on the choice of $\mu$ in the viscosity law, are:
	\small
	\begin{flalign*}
		\rho=14.44\,\,
		k_0^{-3/5}
		\xi^{3/10}
		[\alpha_0m]^{-7/10}
		[\dot{m}J(r)]^{2/5}
		r^{-33/20}
		(1-\tilde{f})^{-3/10}&&
	\end{flalign*}\begin{flalign*}
		T=8.01\times 10^{8}\,\,
		k_0^{-4/15}
		\xi^{-1/5}
		[\alpha_0m]^{-1/5}
		[\dot{m}J(r)]^{2/5}
		r^{-9/10}
		(1-\tilde{f})^{1/5}&&
	\end{flalign*}\begin{flalign*}
		h=1.72\times 10^{-2}\,\,
		k_0^{-7/15}
		\xi^{-1/10}
		[\alpha_0m]^{-1/10}
		[\dot{m}J(r)]^{1/5}
		r^{21/20}
		(1-\tilde{f})^{1/10}&&
	\end{flalign*}\begin{flalign*}
		P=1.91\times 10^{8}\,\,
		k_0^{-13/15}
		\xi^{1/10}
		[\alpha_0m]^{-9/10}
		[\dot{m}J(r)]^{4/5}
		r^{-51/20}
		(1-\tilde{f})^{-1/10}&&
	\end{flalign*}\begin{flalign}
		\label{eq:system_gas}
		\frac{4\alpha_0^2-k_1^4f^4}{k_1^4f^4}=5.41\times 10^{2}\,
		k_0^{-1/5}
		\xi^{-9/10}
		[\alpha_0m]^{1/10}
		[\dot{m}J(r)]^{4/5}
		r^{-21/20}
		(1-\tilde{f})^{9/10} &&
	\end{flalign}
	\normalsize
	
The value of $\xi$ can be obtained by studying the continuity of all the above quantities at the boundary between the radiation pressure- to the gas pressure-dominated regions. It corresponds to $\xi\simeq1.00 k_0^{-1/3}$. 

\begin{figure}[tb]
	\centering
	\includegraphics[width=0.627\columnwidth]{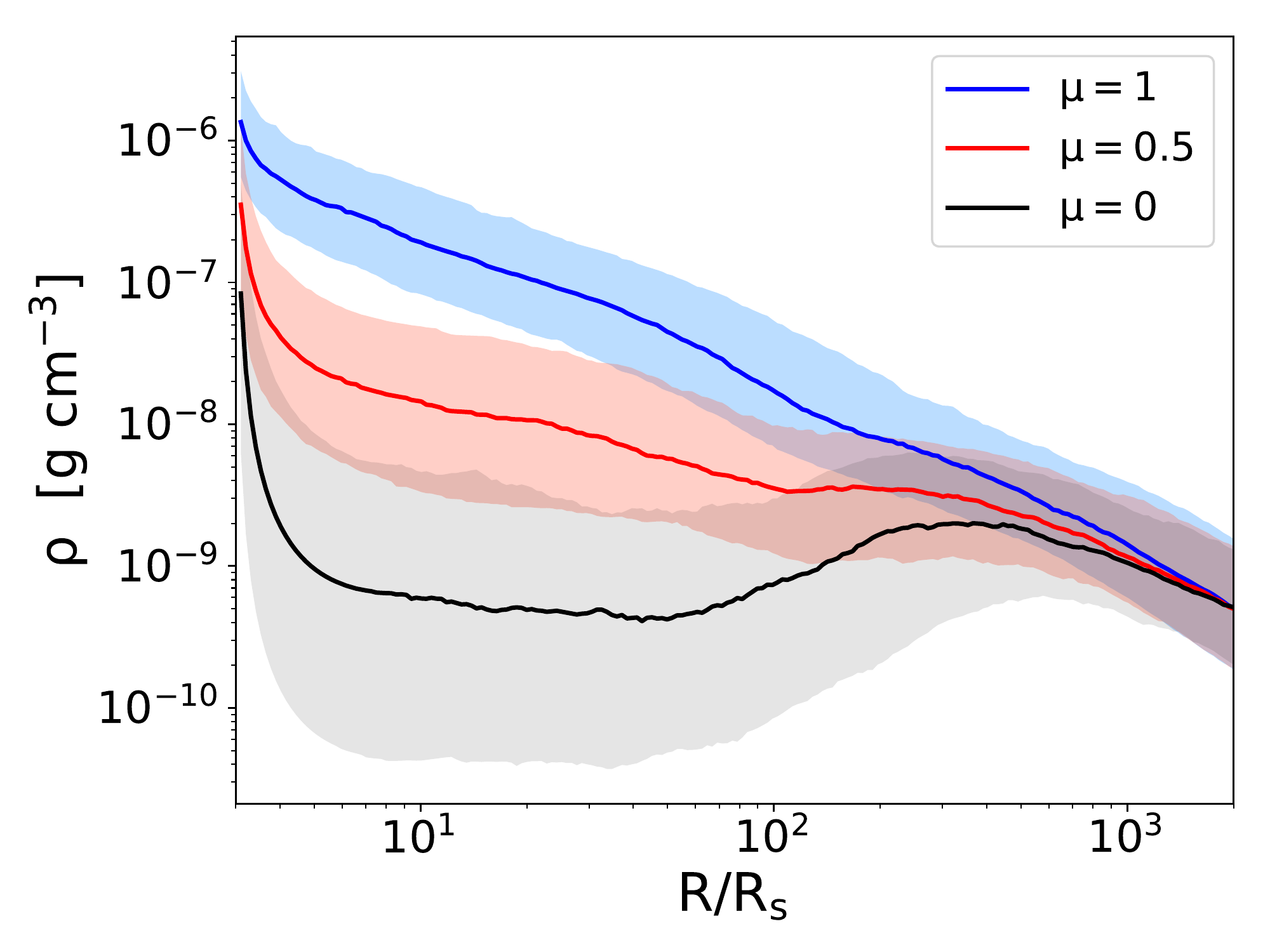} \\
	\includegraphics[width=0.622\columnwidth]{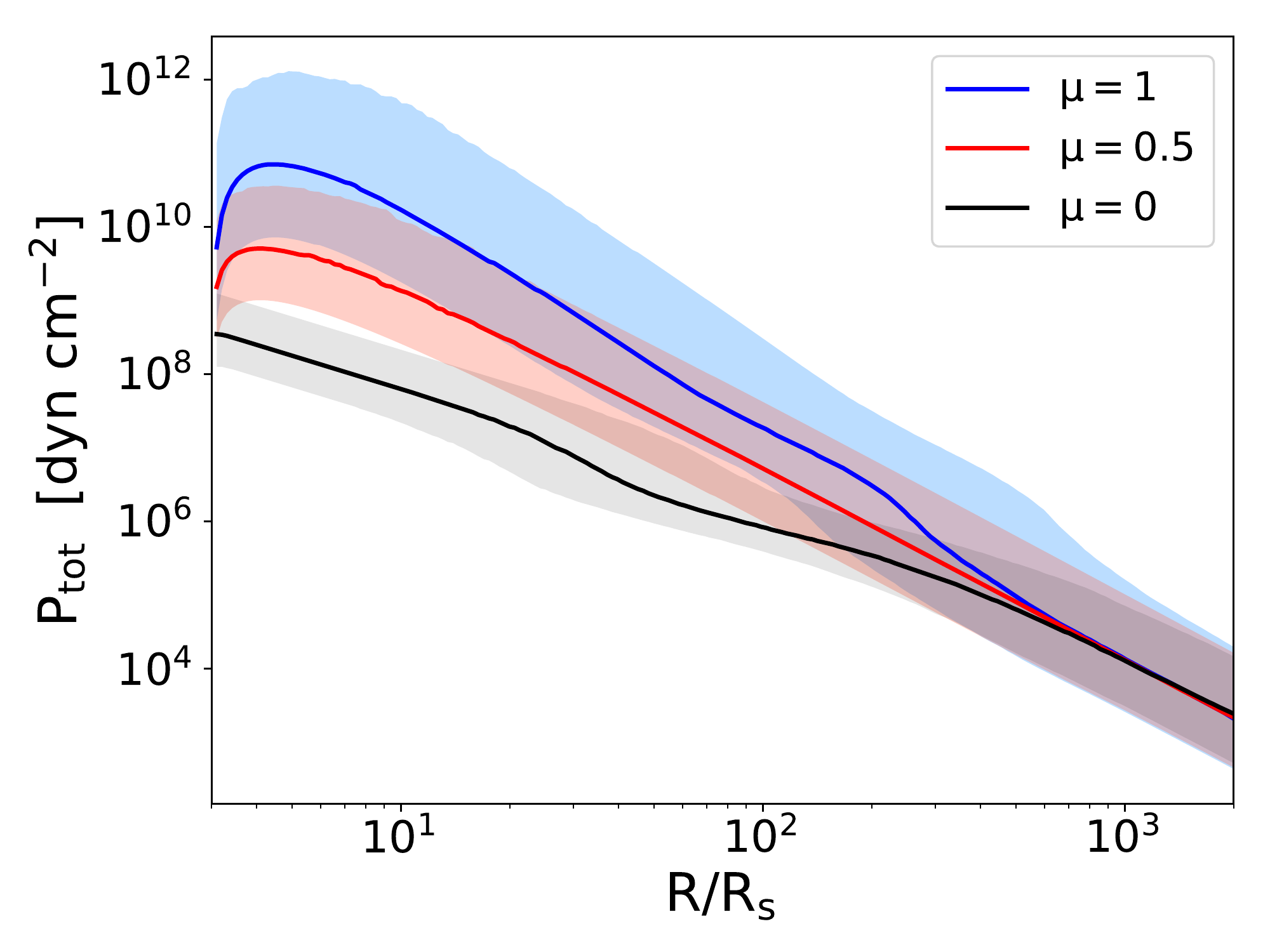} \\
	\includegraphics[width=0.605\columnwidth]{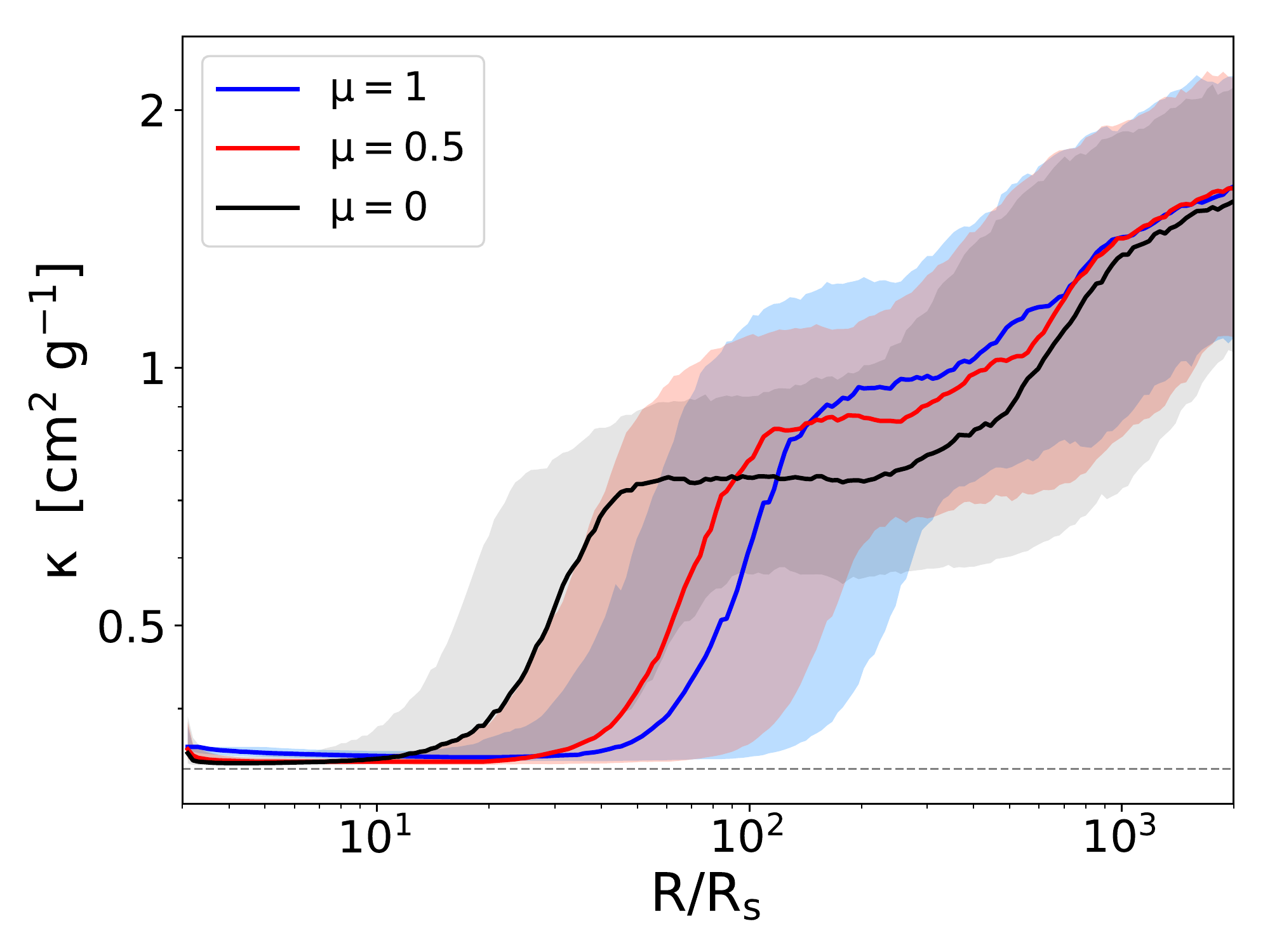}\\
	\includegraphics[width=0.625\columnwidth]{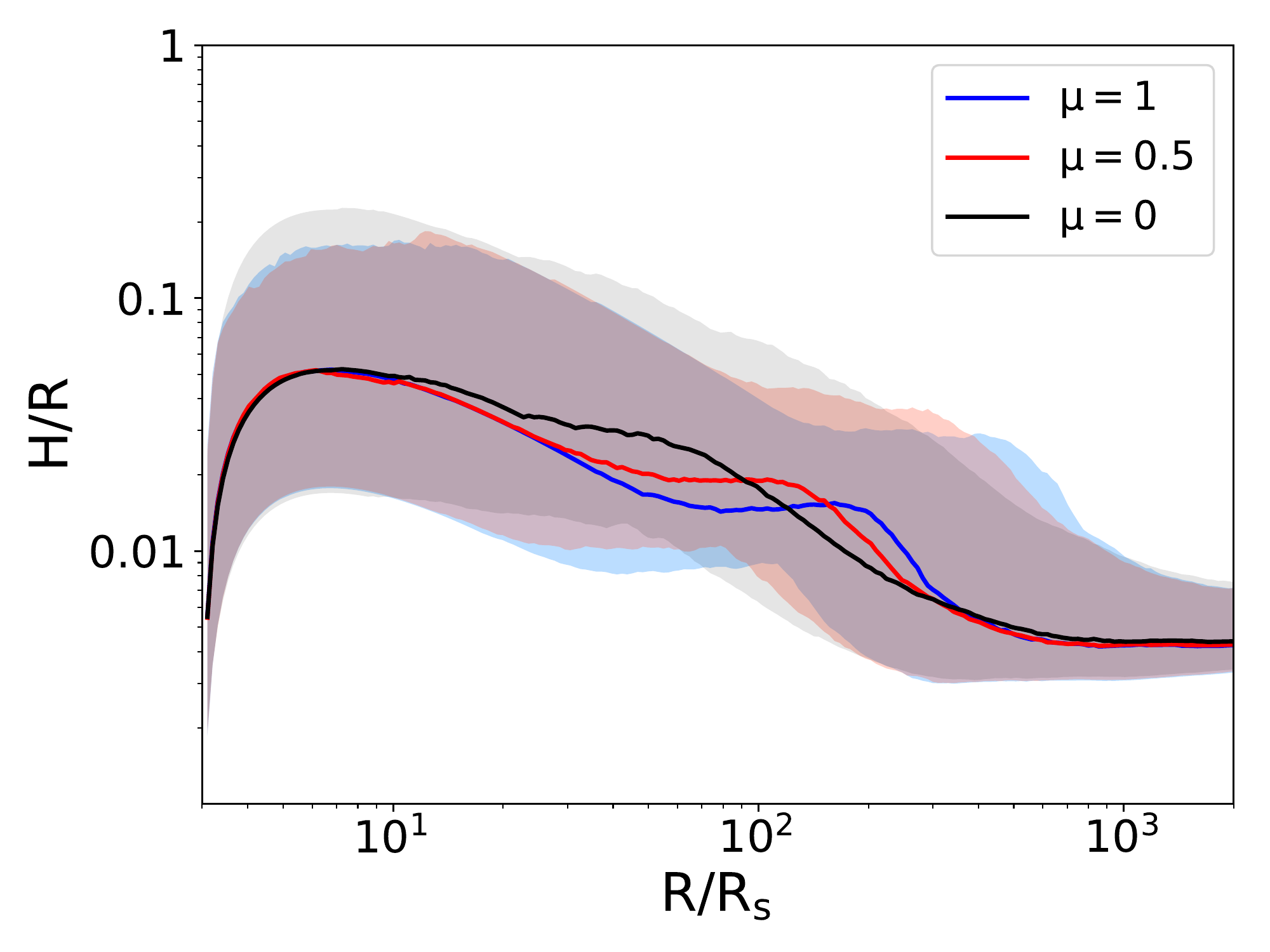} \\
	\includegraphics[width=0.61\columnwidth]{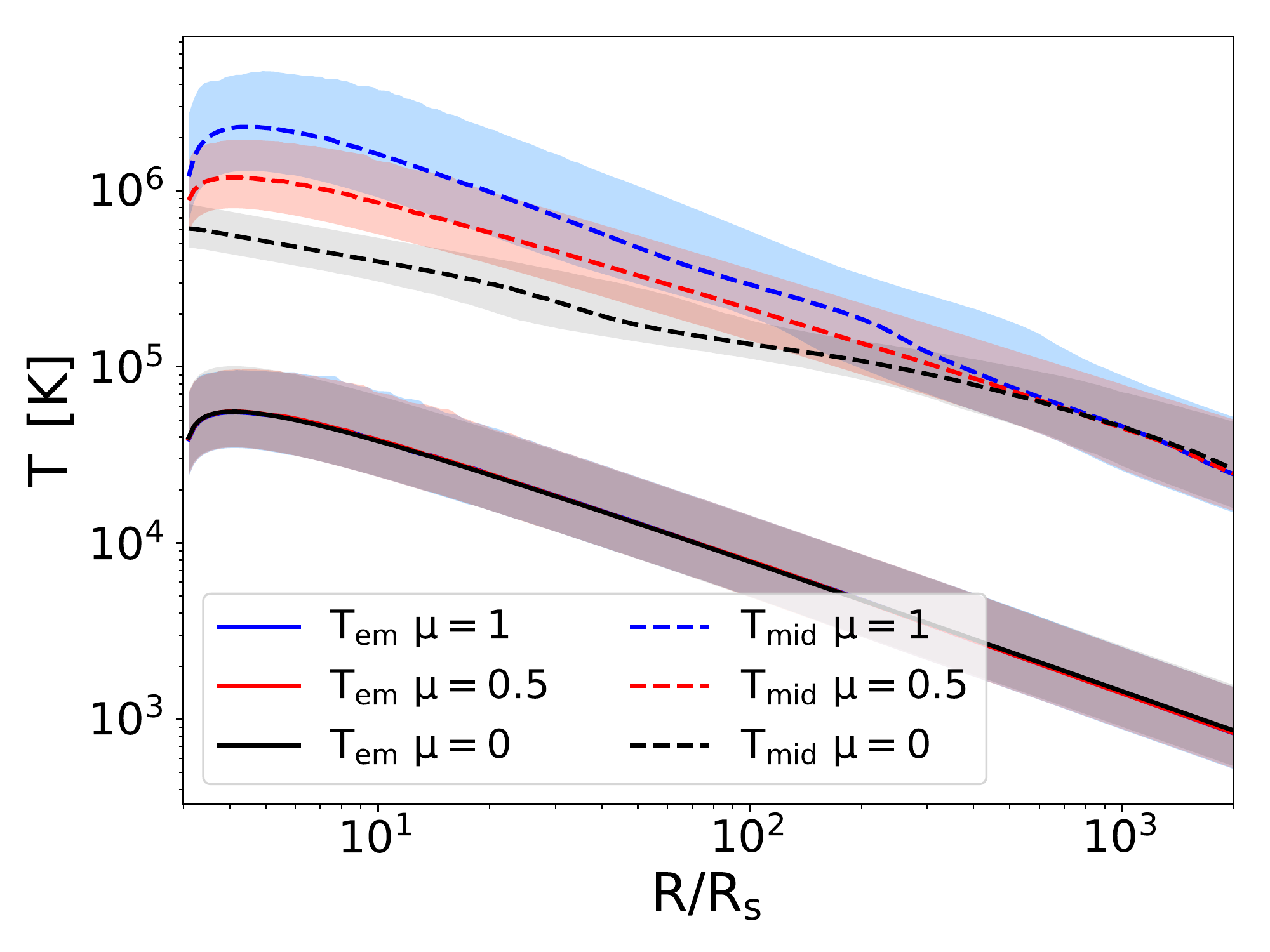}\\
	\caption{Same as Fig.~\ref{fig:f_Lx_profiles}, with radial profiles for the mid-plane $\rho$ (top left), $P_{tot}$ (gas plus radiation, top central), $\kappa$ (top right), $h/r$ (bottom left) and $T$ (both mid-plane and surface, bottom right).}
	\label{fig:other_profiles}
\end{figure}

\begin{figure}[tb]
	\centering
	\includegraphics[width=0.62\columnwidth]{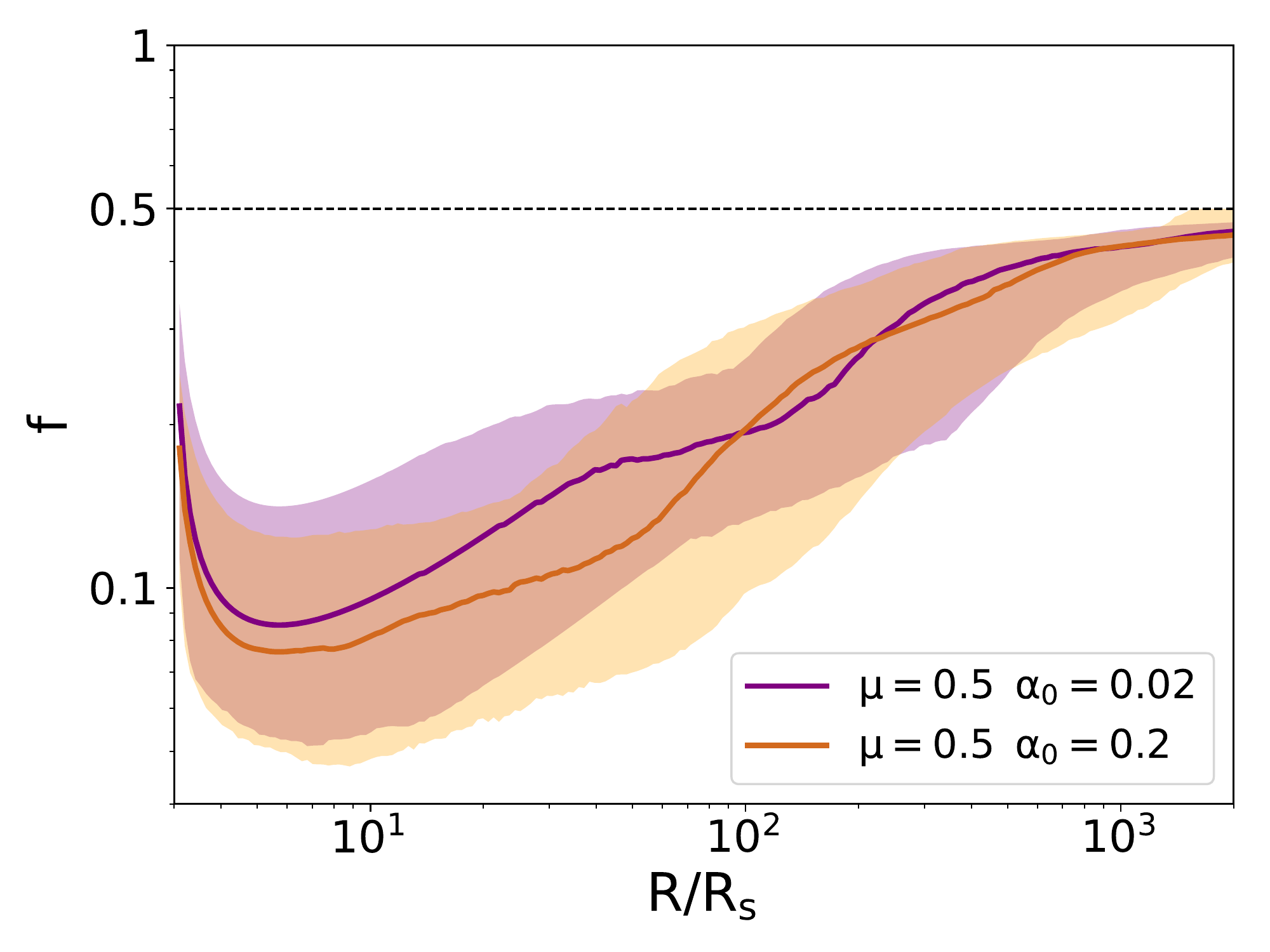}
	\includegraphics[width=0.64\columnwidth]{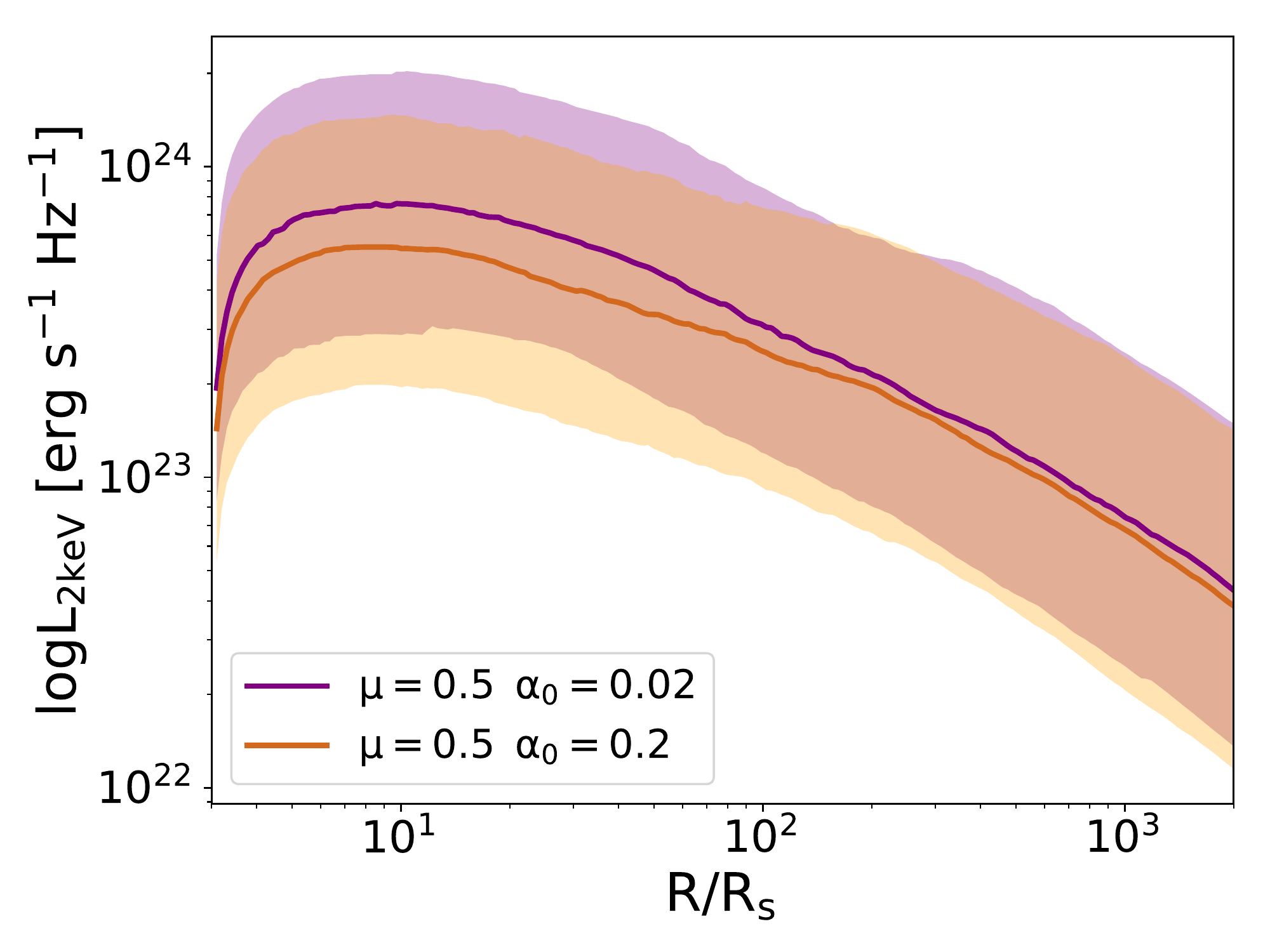}
	\caption{Same as in Fig.~\ref{fig:f_Lx_profiles} and~\ref{fig:other_profiles}. Here, we highlight the (minor) influence on varying $\alpha_0$ from 0.02 (purple) to 0.2 (orange) in $f$- and $L_{2keV}$-profiles.}
	\label{fig:alpha_diff_minor}
\end{figure}

In Fig.~\ref{fig:other_profiles} we report examples of radial profiles for $\rho$, $P_{tot}$, $\kappa$, $h/r$, $T_{mid}$ and $T_{eff}$. Similar examples for $f$ and $L_{2keV}$ are shown in Fig.~\ref{fig:f_Lx_profiles}. Once $f_{max}$ is fixed, the dominant variance among the models is given by the choice of the viscosity law ($\mu$), while $\alpha_0$ plays a minor role. This is shown in Fig.~\ref{fig:alpha_diff_minor}, where profiles for $f$ and $L_{2keV}$ show little difference in varying $\alpha_0$ from 0.02 to 0.2.

\section{The reference AGN sample}
\label{sec:Lx_Luv_XXL}

\begin{figure}[tb]
	\centering
	\includegraphics[width=0.87\columnwidth]{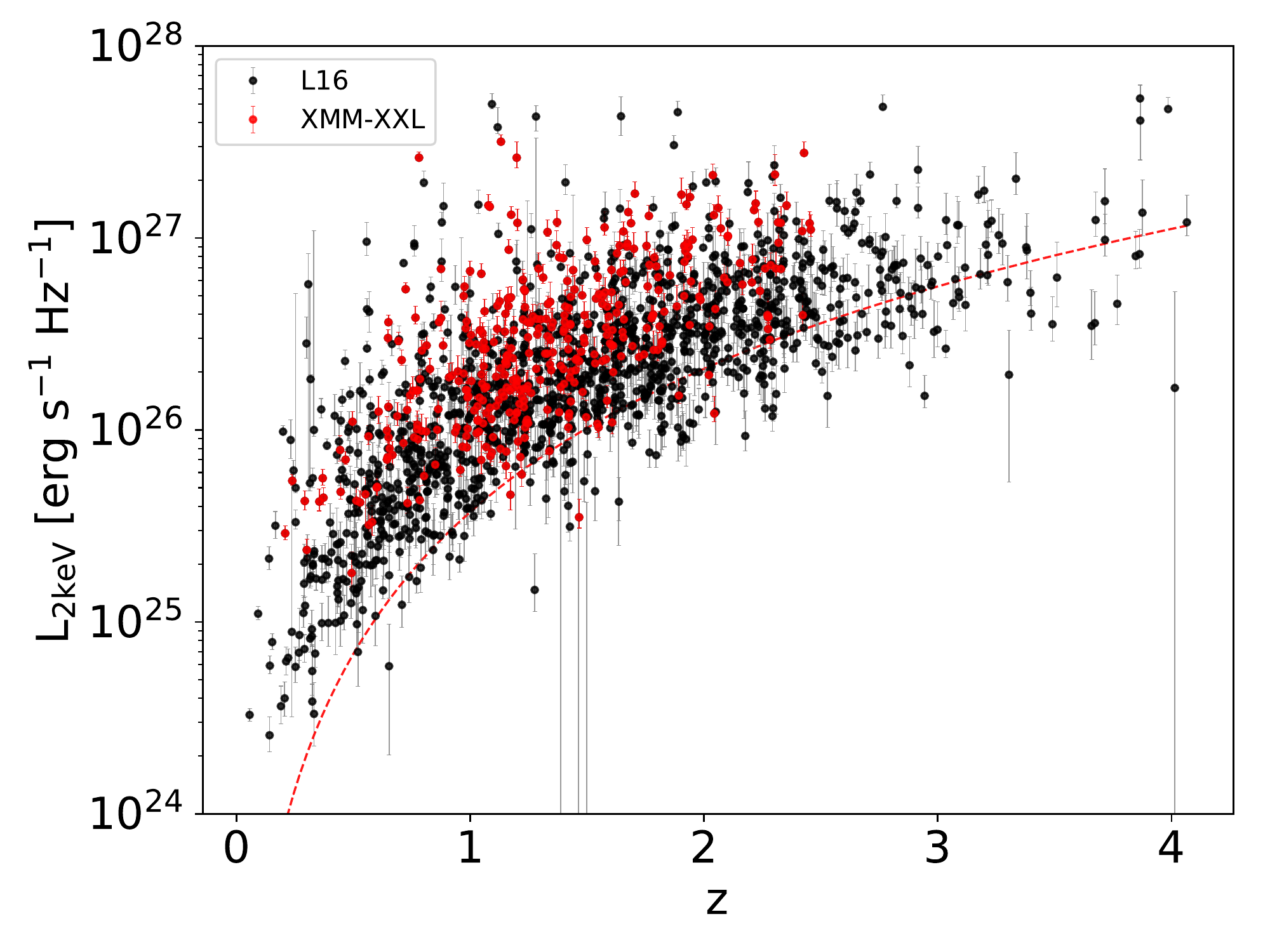} 
	\\
	\includegraphics[width=0.87\columnwidth]{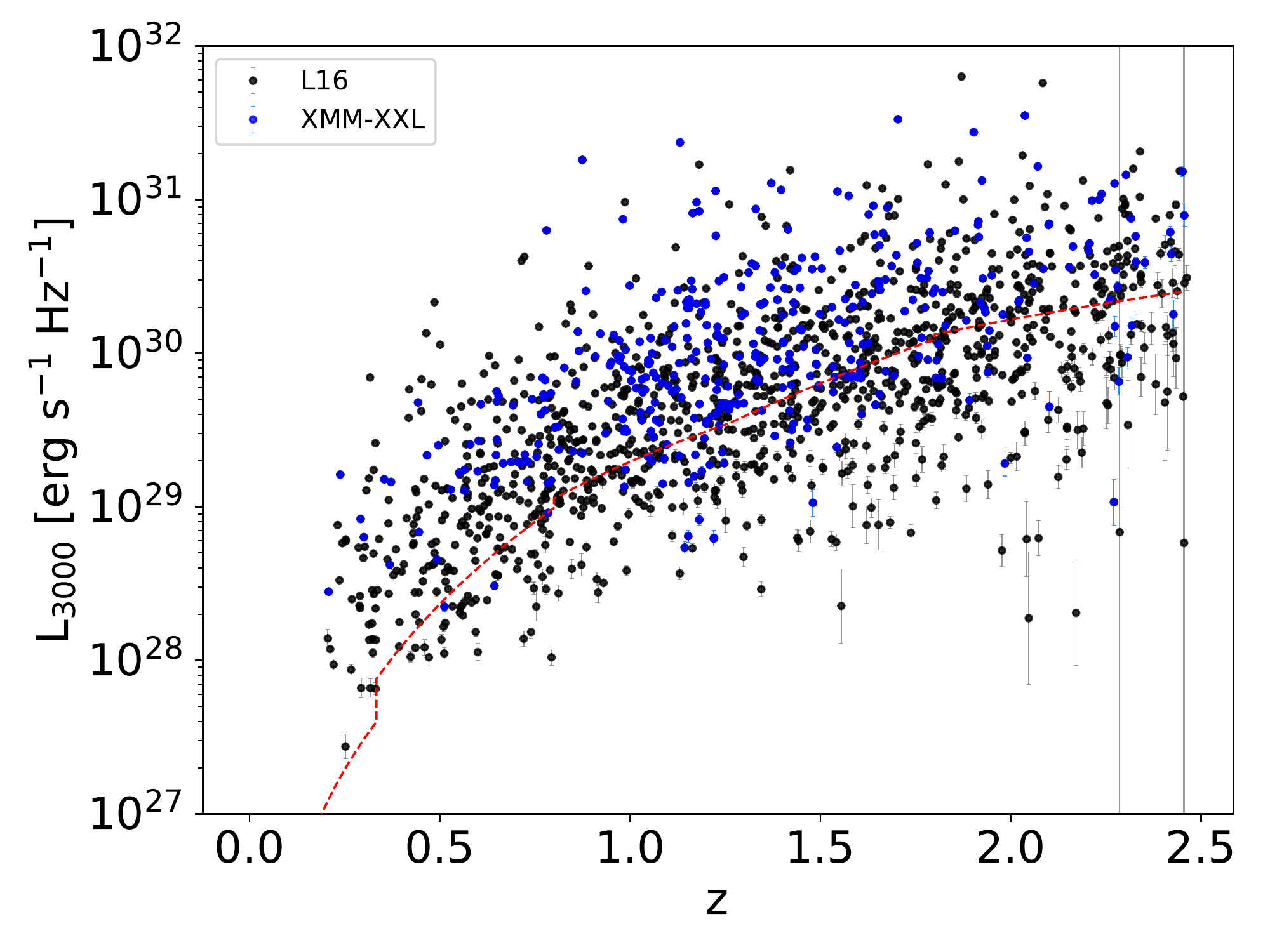}
	\caption{Distribution of $L_{2keV}$ (top panel) and $L_{3000\AA}$ (bottom panel) in the luminosity-redshift plane of the 379 sources of our XMM-XXL sample (red and blue respectively), with respect to the parent sample of BLAGN from \citetalias{Liu_zhu+2016:XMM-XXL} (black). The dashed red lines broadly represent the sensitivity of  the survey at the related frequency (see the text for a description).}
	\label{fig:XMM-XXL_sensitivity}
\end{figure}

For the source-by-source modeling of XMM-XXL we used the 379 sources obtained following the methodology outlined in Section~\ref{sec:method}. In Fig.~\ref{fig:XMM-XXL_sensitivity} we show the distribution of $L_{2keV}$ (top panel) and $L_{3000\AA}$ (bottom panel) in the luminosity-redshift plane of the 379 sources (red and blue respectively), with respect to the parent sample of BLAGN from \citetalias{Liu_zhu+2016:XMM-XXL} (black). The $L_X-L_{UV}$ slope of this reference sample is $0.54\pm0.02$, from Eq.~\ref{eq:best_fit2D}. This is incompatibly flatter than the values quoted in the recent literature, namely $0.64\pm0.02$ \citep{Lusso&Risaliti2016:LxLuvtight} or $0.63\pm0.02$ (\citetalias{Lusso&Risaliti2017:toymodel}). The cleaning criteria applied in Section~\ref{sec:sample} were aimed to exclude low-quality data and to be consistent with the model, while in the above-mentioned literature the possible biases of flux-limited samples were treated carefully in order to reliably use quasars for cosmology \citep{Risaliti&Lusso2018:cosmo2}.

\begin{figure}[tb]
	\centering
	\includegraphics[width=0.99\columnwidth]{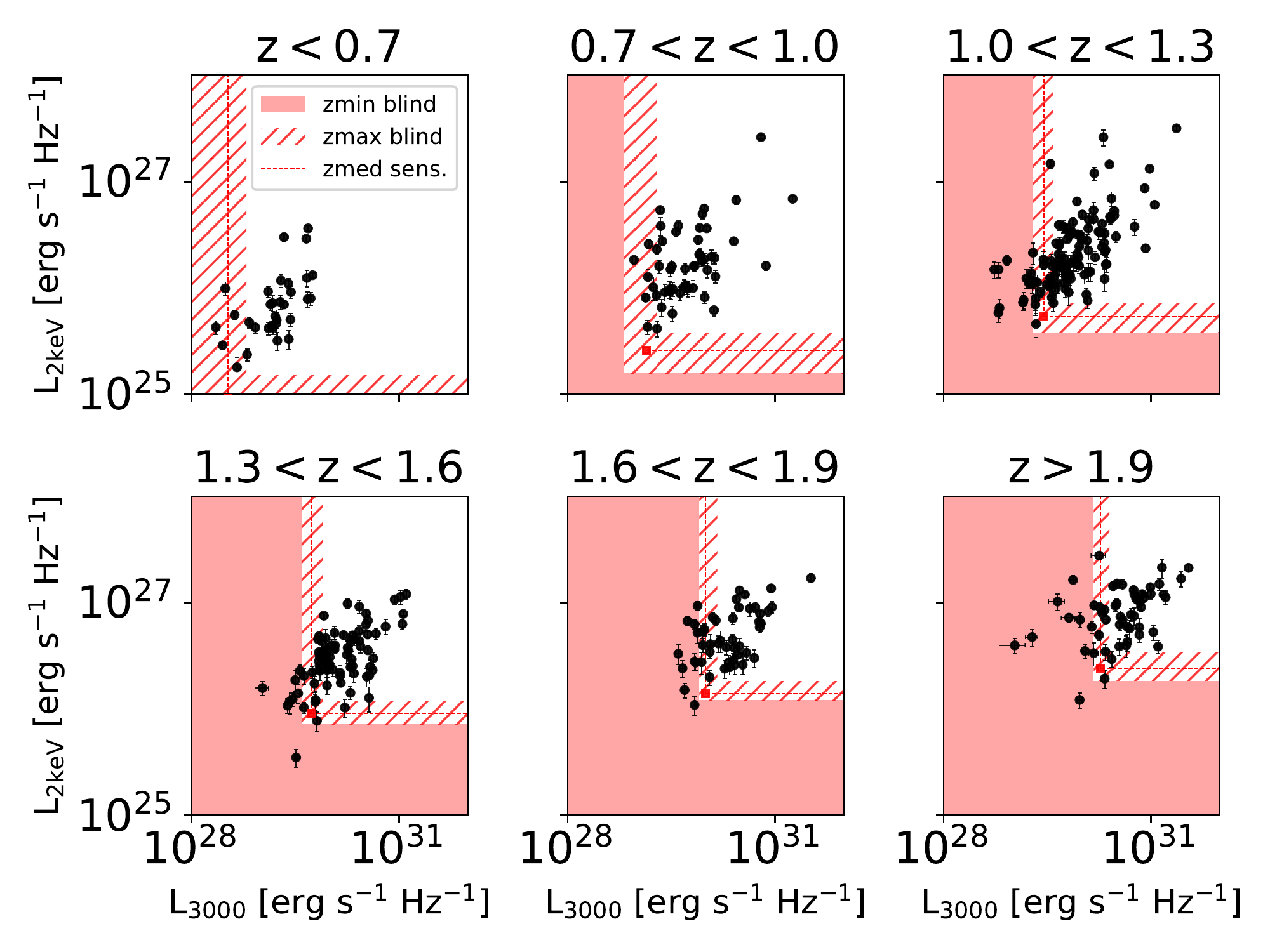}
	\caption{$L_X-L_{UV}$ relation in the redshift bins reported in the sub-titles. The sensitivity surfaces at the minimum, median and maximum redshift of the bin are represented in red with a full area, a dashed line and a shaded area respectively. These surfaces are obtained from the sensitivity lines in Fig.~\ref{fig:XMM-XXL_sensitivity} at the above-mentioned redshifts. The sources above the shaded sensitivity area in each $z$-bin give the cleanest XMM-XXL sample. 
	}
	\label{fig:XMM-XXL_zbins}
\end{figure}

We investigated whether this inconsistency in the slope would be bridged restricting the analysis to the brightest objects at all redshifts with a very crude and conservative selection. From the sensitivity curve of the XXL-N survey in the $0.5-10\,$keV band at half of the survey area (\citetalias{Liu_zhu+2016:XMM-XXL}, their Fig. 3) we obtained the flux limit in that energy band. Then, we interpolated the flux limit at 2\,keV using the mean photon index of the sample, obtaining the sensitivity curve shown in red in the top panel of Fig.~\ref{fig:XMM-XXL_sensitivity}. In \citet{Menzel+2016:XXM-XXL_boss} a cut at $r<22.5$\,mag was applied. We converted this magnitude limit in a luminosity sensitivity only within $0.80\lesssim z\lesssim 1.27$, for which $3000\AA$ was actually detected in the $r$ band. For different redshifts, we first computed a redshift dependent color correction for the other bands ($u$,$g$,$i$ and $z$) performing a linear regression on the difference with the $r$-band magnitude. This provided a magnitude limit for $L_{3000\AA}$ at all redshifts, consistently with the band in which that wavelength was actually detected, from which we obtained the related sensitivity line in the bottom panel of Fig.~\ref{fig:XMM-XXL_sensitivity}. We then divided XMM-XXL in six redshift bins, making sure to have at least 30 counts per bin. For each bin, we excluded all the sources below the limits given by the sensitivity curves on both axis, evaluated at the maximum $z$ of the bin to be conservative (Fig.~\ref{fig:XMM-XXL_zbins}). The resulting cleanest subsample reaches accordance with the recent literature of the $L_X-L_{UV}$, with a slope of $0.59\pm0.03$. 

\end{appendix}
			
\end{document}